\UseRawInputEncoding
\documentclass[journal=jacsat,manuscript=article]{achemso}

\newcommand{\CHANGE}[1]{#1}
\newcommand{\REMOVE}[1]{}

\usepackage[version=3]{mhchem} 
\usepackage{siunitx}
\usepackage{booktabs}
\usepackage[labelfont=bf]{caption}
\usepackage{doi}
\usepackage{xcolor} 

\SectionNumbersOn
\newcommand*\mycommand[1]{\texttt{\emph{#1}}}
\newcommand*{\degree}{^\circ}

\author{M. A. Basyooni-M. Kabatas}
\email{m.kabatas@tudelft.nl}
\affiliation[PME]{Department of Precision and Microsystems Engineering, Delft University of Technology, Mekelweg 2, 2628 CD Delft, The Netherlands}

\author{Tao Shen}
\affiliation[PME]{Department of Microelectronics, Delft University of Technology, Delft, the Netherlands}

\author{Ka\"i Betlem}
\affiliation[PME]{Department of Precision and Microsystems Engineering, Delft University of Technology, Mekelweg 2, 2628 CD Delft, The Netherlands}

\author{Chunyu Huang}
\affiliation[TNW]{Chemical Engineering Department, Delft University of Technology, 2629 HZ Delft, The Netherlands}

\author{Monique A. van der Veen}
\affiliation[TNW]{Chemical Engineering Department, Delft University of Technology, 2629 HZ Delft, The Netherlands}

\author{Frans Widdershoven}
\affiliation[EWI]{Department of Microelectronics, Delft University of Technology, Delft, the Netherlands}
\alsoaffiliation[NXP]{NXP Semiconductors, Technology \& Operations/CTO Office, High Tech Campus 46, Eindhoven 5656 AE, The Netherlands}

\author{Murali K. Ghatkesar}
\affiliation[PME]{Department of Precision and Microsystems Engineering, Delft University of Technology, Mekelweg 2, 2628 CD Delft, The Netherlands}

\author{Peter G. Steeneken}
\email{p.g.steeneken@tudelft.nl}
\affiliation[PME]{Department of Precision and Microsystems Engineering, Delft University of Technology, Mekelweg 2, 2628 CD Delft, The Netherlands}

\title{Capacitive Pixelated CMOS Electronic Nose}

\begin{document}

\begin{abstract}
Although some of the human senses can nowadays be replaced by low-cost electronic sensors such as microphones and image sensors, a compact low-cost electronic nose (E-nose) remains elusive. In this work, an E-nose is presented that can capacitively detect volatile organic compounds (VOCs). The E-nose consists of an array of 1024 capacitive microelectrodes on a complementary metal-oxide-semiconductor (CMOS) chip, functionalized by inkjet printing. The pixels are coated with a UV-curable ink and metal-organic frameworks (MOFs: ZIF-8, MIL-101(Cr), MIL-140A) to create chemically diverse microdomains that generate gas-specific response patterns through adsorption-driven dielectric loading. ZIF-8 exhibits the highest response to 2-butanone, whereas the UV-curable layer responds most strongly to toluene; both show low cross-sensitivity to water vapor, enabling operation under humid conditions. After calibration in pure gases, reproducible responses to controlled binary mixtures of toluene and 2-butanone are observed. The device operates at low power, combines a large 1024-pixel array with CMOS integration, and offers application-specific functionalization by inkjet printing, providing both low cost and versatility. By further extending the range of functionalization materials, the E-nose can be applied to analyze a wide variety of gases, with potential applications in safety monitoring, health, agriculture, and robotics.

\end{abstract}



\setcounter{section}{0}
\section{Introduction}

Identification of volatile organic compounds (VOCs) is important for food quality control, health monitoring, and pest management.\cite{liu2023volatile} Various VOC detection techniques exist, including infrared (IR) spectrometers,\cite{xia2024advancements} laser absorption spectroscopy (LAS),\cite{mhanna2023laser} and gas chromatography coupled with mass spectrometry (GC-MS).\cite{shuaibu2024traceability} \CHANGE{However, these techniques require instruments that are not easily portable, consume substantial power, and are relatively expensive. Since many applications require detectors that can continuously monitor the gas composition in situ, for example in early pest management in agriculture, indoor air-quality monitoring, and food spoilage detection, there is a strong need to develop compact, low-cost electronic noses (E-noses).}

\CHANGE{Electronic noses combine arrays of broadly responsive chemical sensors with pattern-recognition algorithms to generate odour fingerprints that enable discrimination of complex VOC mixtures. They have been applied in diverse domains, including breath analysis for medical diagnostics and disease screening,\cite{DiNatale2014,Aina2024,Sachan2025} environmental and indoor air-quality monitoring, food and beverage quality control, and industrial or safety-related VOC surveillance.\cite{Li2024} Recent work has also demonstrated robust chemical classification using advanced chemosensor platforms coupled with machine learning, for example graphene-based sensor arrays.\cite{Pannone2024} In these scenarios, E-noses complement analytical techniques such as GC-MS and spectroscopic methods, which remain the gold standard for detailed laboratory analysis but are less suited to compact, low-power, continuous field deployment.}

Recent studies have set steps towards this goal by demonstrating real-time VOC detection with E-nose sensors that electronically detect gases via property changes in sensing materials, affecting capacitance,\cite{mahmud2021low,dang2024facile,fernandez2021printed,Saif2025,Darwish2025} resistance,\cite{yao2016mof} or luminescence\cite{shustova2013selective} of dedicated functionalization materials, which are deposited between the electrodes of the E-nose sensor. Most of these works focus on a single or low number of sensing elements per chip. \CHANGE{In the present work, we deliberately employ a capacitive transducer architecture rather than a resistive one. In our CMOS implementation,\cite{laborde2015real} the metal electrodes are fully covered by a dielectric layer and are therefore not in direct contact with the sensing film. This configuration protects the electrodes from electrochemical attack and contamination, which improves the long-term stability and chemical robustness of the device. Moreover, the capacitive signal originates from adsorption-driven changes in the effective dielectric constant of the functional layer, without any DC current flowing directly through the sensing material. As a result, Joule heating and electrically induced degradation of the sensing film are avoided, leading to more stable and reproducible responses over many exposure cycles.} 

\CHANGE{In typical E-nose concepts, individual vapour sensors are not designed as highly specific ``lock-and-key'' receptors for a single analyte. Instead, each sensor element is broadly responsive to a range of vapours and exhibits a characteristic response pattern across different analytes. By combining the parallel outputs of an array of such partially-specific sensors, an odour fingerprint is obtained that can be used to discriminate and quantify complex gas mixtures after appropriate training.\cite{Persaud1982} This principle was already demonstrated by Persaud and Dodd, who showed that a model electronic nose can reproducibly distinguish a wide variety of odours without requiring highly specific receptors.\cite{Persaud1982} Consequently, the role of a large sensor array is not to provide one unique sensor per gas, but to span a multidimensional response space in which different VOCs and mixtures map to separable patterns.}

\CHANGE{As sensing layers, we employ MOFs, a class of porous crystalline materials composed of metal nodes and organic linkers that can offer high surface areas, tunable pore structures, and chemical robustness, making them attractive for VOC sensing. In this work, we focus on three representative MOFs: ZIF-8, MIL-101(Cr), and MIL-140A. ZIF-8 is a zeolitic imidazolate framework with small hydrophobic cages and narrow windows; MIL-101(Cr) features large three-dimensional cages with coordinatively
unsaturated metal sites; and MIL-140A is a one-dimensional channel system with Zr–O backbones. These materials were selected because they combine distinct pore architectures and hydrophilicity with good thermal and water stability, allowing us to study how different MOF structures influence VOC adsorption and sensor response. The detailed structural and adsorption characteristics of these MOFs are summarized in the Supporting Information.}

To realize such an E-nose with multiple sensor elements requires addressing two key challenges:
\begin{enumerate}
    \item Realizing a large number of sensor elements that can be individually read out on a small area with low-cost electronics.
    \item Providing multiple different functionalization materials in the sensor area.
\end{enumerate}
Here, we present a pixelated capacitive CMOS electronic nose with on-chip readout and inkjet-printed MOF-based and polymer-based functional layers. After calibration in pure gases, the E-nose yields reproducible, gas-specific response patterns and supports concentration estimation for single-component exposures using linear or quadratic calibration models. Responses to binary gas mixtures are reported qualitatively.

\section{Experimental} 
\label{sec:experimental}
\subsection{Nanoparticle synthesis and characterization}

\subsubsection{Chemicals}
Three types of metal-organic framework (MOF) nanoparticles were synthesized in powder form, of which ZIF-8 was found to be the most suitable, and was therefore used for the final E-Nose experiment. The synthesis of MOF nanoparticles was carried out using the following high-purity chemical reagents. Zinc nitrate hexahydrate (98\%, Sigma-Aldrich) and 2-methylimidazole (98\%, TCL) were used as precursors for ZIF-8 synthesis. For the preparation of MIL-140A, N,N-Dimethylformamide (DMF) (\(\geq 99.9\%\), Sigma-Aldrich), terephthalic acid (98\%, Merck Sigma), and zirconium chloride (99.5\%, Thermo Fisher Scientific) were employed. Chromium chloride hexahydrate (96\%, Merck Sigma) and terephthalic acid (98\%, Merck Sigma) were utilized for MIL-101(Cr) synthesis.
Additionally, methanol (\(\geq 99.9\%\), Honeywell), and ethanol (99.5\%, Thermo Fisher) were utilized in the respective synthesis processes. All aqueous solutions were prepared using ultrafiltrated deionized (DI) water with a resistivity of 18.0 MΩ·cm, obtained from an LWTN Genie A system (Laboratorium Water Technologie Nederland).

\subsubsection{Synthesis of ZIF-8}
ZIF-8 was synthesized following a previously reported method \cite{pan2011rapid}, with slight modifications. First, 1.17 g of Zn(NO$_3$)$_2\cdot$6H$_2$O was dissolved in 8 mL of deionized (DI) water. Separately, 22.7 g of 2-methylimidazole was dissolved in 80 mL of DI water. The two precursor solutions were mixed and stirred at room temperature (~22°C) for 5 minutes. The resulting product was observed to immediately become a milky suspension. The precipitate was collected by centrifugation, washed three times with methanol at 15,000 Relative Centrifugal Force (RCF) for 30 minutes, and finally dried in a vacuum oven at 65°C overnight.

\subsubsection{Synthesis of MIL-101(Cr)}
MIL-101(Cr) was synthesized hydrothermally using 1.26 mmol of CrCl$_3\cdot$6H$_2$O and 1.26 mmol of terephthalic acid. These compounds were combined in a 40 mL autoclave and dispersed in 20 mL of deionized (DI) water using 10 minutes of sonication. The autoclave was then sealed and placed in an oven at 180°C for 48 hours. After cooling to room temperature (RT), the autoclave was opened, and the precipitate was manually stirred using a disposable spatula to redisperse the particles. The product was transferred to 50 mL centrifuge tubes and centrifuged at 15,000 RCF for 30 minutes. The resulting supernatant was removed, and further washing was performed to eliminate residual reactants and unreacted precursors: the sample was resuspended in 30 mL of N,N-Dimethylformamide (DMF) using 10 minutes of sonication. The mixture was poured back into the autoclave, resealed, and heated at 60°C for 3 hours. After cooling to RT, the mixture was transferred back into centrifuge tubes and centrifuged at 15,000 RCF for 30 minutes. This washing process was repeated twice with DMF and once with ethanol (EtOH) to ensure thorough purification. After the final washing step with ethanol, the residue was dried overnight in a vacuum oven at RT. The dried as-synthesized samples were weighed and collected in 5 mL glass sample vials. The final activation step was carried out by drying the powder in a vacuum oven at 150°C for 8 hours under vacuum, ensuring complete removal of residual solvents and achieving full activation of the material.

\subsubsection{Synthesis of MIL-140A}
MIL-140A was synthesized by dissolving 1.75 mmol of ZrCl$_4$ in 10 mL of DMF and 1.75 mmol of terephthalic acid in 10 mL of DMF. Both solutions were ultrasonicated for 20 minutes, then combined and sonicated for an additional 5 minutes. The resulting solution was transferred into a 40 mL PTFE-lined autoclave and heated at 140°C for 24 hours. After synthesis, the product was washed three times with DMF. A solvent-exchange step was performed by soaking the powder in methanol for 24 hours, after which the methanol was discarded. The powder was then dried in a vacuum oven at RT for 24 hours, followed by an additional drying step at 80°C for 24 hours.

\CHANGE{The structural and chemical characteristics of ZIF-8, MIL-140A, and MIL-101(Cr), including their formulas and pore dimensions, are summarized in Table~\ref{tab:mof_structure} and in the Supporting Information, Section~S2, Table~S1.}

\subsubsection{Characterization}
Powder X-ray diffraction (PXRD) was performed using a Bruker D8 Advanced diffractometer with a Cu K$_\alpha$ source ($\lambda = 1.5418$ \r{A}). The measurements were conducted over a 2$\theta$ range of 3-70°, allowing for phase identification and crystallinity assessment. Thermogravimetric Analysis (TGA) was carried out using a Mettler Toledo 1600 TGA/SDTA851 under synthetic air conditions within a 30-800°C temperature range, with a 100 mL/min gas flow rate. The pre-treatment involved heating at 30°C for 30 minutes, followed by a temperature ramp of 10°C/min up to 800°C, providing insight into the thermal stability of the materials. 
The morphological characterization of the MOF particles was conducted using Scanning Electron Microscopy (SEM) on a Jeol JSM-IT700HR. The imaging parameters used for SEM analysis are detailed in the corresponding figure. Energy Dispersive X-ray (EDX) analysis was performed using a Thermo Fisher Helios G4 UXe PFIB dual beam system. The EDX detector was an Ametek EDAX Octane Elite 65 mm², operated with the EDAX APEX EDS software suite, with acceleration voltages of 2 and 5 kV, and the beam current was 50 pA. Optical microscopy images of the chip and inks were captured using a Keyence VHX-600 digital microscope. 
MOF crystal structure images were generated by VESTA software, which is a three-dimensional visualization system for electronic and structural analysis.

Gas adsorption studies were carried out to evaluate the porosity and surface area of the MOFs. N$_2$ adsorption isotherms were obtained using a Micromeritics Tristar II at 77 K. Before the N$_2$ adsorption measurements, ZIF-8 was degassed under N$_2$ gas flow at 120°C for 16 hours, while MIL-101(Cr) and MIL-140A were degassed under the same conditions but at 150°C for 16 hours to ensure complete removal of adsorbed species. The BET surface area was calculated following Rouquerol criteria using the BETSI program~\cite{ROUQUEROL200749,osterrieth2022reproducible}.  To study CO$_2$ uptake, CO$_2$ adsorption isotherms were recorded using a Micromeritics Tristar II at 298 K across a pressure range of 0-120 kPa. Before conducting CO$_2$ adsorption, ZIF-8 was degassed at 120°C for 3 hours under N$_2$ flow, whereas MIL-101(Cr) and MIL-140A were degassed at 150°C for 3 hours under N$_2$ gas flow. Water adsorption behavior was evaluated using a Micromeritics 3Flex Surface and Catalyst Characterization system at 293 K. Before the H$_2$O adsorption measurements, ZIF-8 was degassed in a vacuum degasser at 120°C for 16 hours, whereas MIL-101(Cr) and MIL-140A were degassed in a vacuum degasser at 150°C for 16 hours to eliminate any residual moisture.
\CHANGE{All adsorption isotherms have been converted to the AIF format and provided as supplementary material \cite{evans2021universal}.}

\subsection{Functionalization by inkjet printing}
\subsubsection{Ink solvents}

High-purity solvents were used in the ink formulation experiments. Ethanol (\(\geq 99.5\%\), Thermo Fisher) was used to prepare the inks, selected for its compatibility with the MOF materials and its effectiveness in dissolving the ink components. 2-propanol (\(\geq 99.7\%\), Sigma-Aldrich) was used for surface treatment of the CMOS chip.

\subsubsection{Ink formulation}
A commercially available Canon IJC257 UV-curable ink, referred to as UV ink, is used in this study. The ink is a mixture of multifunctional acrylate monomers/oligomers; mainly 3-methyl-1,5-pentanediol diacrylate, neopentyl-glycol diacrylate, 2-ethylhexyl acrylate, and trace acrylic acid, together with a benzophenone-type photoinitiator and proprietary additives. After UV exposure, the acrylate groups cross-link into a densely packed polyester/urethane network whose matrix is largely hydrophobic but still contains ester carbonyls that act as weak dipole sites. In this section, we explain how the MOF-based inks were prepared. To prepare the ZIF-8 ink, 0.5 g of dried ZIF-8 nanopowder was mixed with 2 mL of a 50:50 (v/v) water-ethanol solution. The mixture was sonicated for 20 minutes at an intensity of 270 W using a sonication probe in an ice bath to prevent overheating. The solution was then filtered using a 0.1 $\mu$m syringe filter to remove larger particles. A similar preparation method was applied for MIL-101(Cr) and MIL-140A. MIL-140A required a slightly longer sonication time of 60 min to achieve complete dispersion.
Next, 2\% (v/v) of the UV ink was added to the ink solution, followed by brief sonication to ensure thorough mixing using a Nanografi Ultrasonic Probe Sonicator at 30\% power intensity in an ice bath to minimize mechanical and thermal damage to the MOFs. This step was observed to enhance the adhesion of MOF particles to the CMOS chip.

The final mixture consisted of 2\% UV ink and 98\% MOF suspension, a ratio selected to preserve accessibility to the MOF’s high porosity, while ensuring compatibility with the inkjet printing process. This composition also helped maintain the MOF ink on the chip surface for an extended period and prevented MOF particles from detaching after drying. After printing on Si wafers or CMOS chips, an annealing process was performed using Memmert UN30 at 100°C for 30 minutes in air to eliminate residual solvents and improve film stability. The selection of low-boiling-point solvents, such as water and ethanol, in this study was intentional. This approach helps to avoid high-temperature annealing of the CMOS chip for extended durations, thereby preventing potential damage to wire bonding and electronic components. Moreover, higher annealing temperatures require a higher UV ink polymer content to ensure MOF stability on the CMOS chip. However, this increase in polymer content could negatively impact the porosity of the MOF-based ink and, consequently, its VOC selectivity.

\subsubsection{Substrate pretreatment}
Pristine (100) silicon or silicon/silicon dioxide (Si/SiO$_2$) wafers with a 300 nm thermal SiO$_2$ layer, a 4-inch diameter, a thickness of 525 $\pm$ 25 $\mu$m, and a resistivity of $10^{3} \text{ to } 10^{5} \ \Omega \cdot \text{cm}$ were obtained from MicroChemicals (Germany). These wafers served for testing the inkjet printing parameters before starting the printing procedure on the CMOS chips. Before inkjet printing, the substrates were cleaned by sonicating them in acetone for 15 minutes, followed by 2-propanol (2IPA) for another 15 minutes. The substrates were then rinsed with deionized (DI) water and dried using an N$_2$ gun. It is important to note that no plasma treatment was applied to increase the surface hydrophilicity, in order to avoid potential electrostatic discharge (ESD) damage or degradation of the chip’s wire bonds during vacuum plasma exposure.

Before inkjet printing on the CMOS chips, the chip surface was rinsed with 2-propanol and dried using a N$_2$ gun to remove any organic residues or dust, thereby ensuring optimal ink adhesion and print quality.

\subsubsection{Inkjet printing setup and parameters}

The PIXDRO LP-50 inkjet printer, equipped with Dimatix Materials Cartridge (DMC)-11610 cartridges, was used for the precise deposition of MOF nanoparticles and UV inks. This printer operates using a piezo-driven drop-on-demand mechanism, allowing fine control over the printing process with high resolution. The DMC printhead consists of 16 individually controlled nozzles, but in this study only a single 10~pL nozzle was activated for each ink.

The inkjet printing process was optimized for uniform deposition of UV and MOF-based inks on the CMOS chip and on Si/SiO\textsubscript{2} wafers. A piezo-actuated nozzle was used for each ink under different voltages and back-pressure conditions to ensure precise ink delivery. The distance between the print head and the substrate was set at 7.7~mm for the CMOS chip and 1~mm for the Si wafer, as these values provided optimal ink spreading and adhesion. The substrates were fixed on an $x$–$y$ stage to prevent movement during printing, and all depositions were performed at room temperature ($\sim$29~\si{\celsius}). After the printing process, UV curing was applied for 5~s using the built-in UV light source of the PIXDRO LP-50 to enhance ink adhesion and stability.

The detailed printing parameters are provided in Table~\ref{table:ink-printing}, which lists the printhead type, voltage settings, print speed, ink pressure, resolution, droplet volume, and pulse-shape characteristics. The pulse-shape voltages and time durations (Idle Before, Fill Ramp, Time Low, Fire Ramp) are crucial factors affecting droplet formation and deposition quality. The UV ink and MOF-based inks (ZIF-8, MIL-101(Cr), and MIL-140A) were all printed using a DMC-11610 printhead, with the driving voltage, back pressure, and waveform timings optimized to generate stable, satellite-free droplets. This printing approach allows for the precise formation of ink patterns with minimal variation in thickness and ensures good repeatability across different printed attempts, as shown in Fig.~\ref{fig:inkjet-patterns}.

\CHANGE{To further control ink deposition during printing, the printhead--substrate distance, print speed, and substrate temperature were kept constant for all prints, and droplet sizes and shapes were first optimized on Si/SiO\textsubscript{2} wafers before printing on the CMOS chip. Optical microscopy and FESEM images (Fig.~S10) and patterned arrays (Fig.~S13) confirm well-defined circular droplets and uniform droplet distributions, ensuring a reproducible deposited volume per droplet. On the CMOS chip, the number and spacing of droplets per functional region are defined in the printing software so that each region receives a controlled number of overlapping droplets, leading to a consistent effective coating thickness over the selected pixel ensembles (see also Fig.~5(b–g) and Fig.~S15).}

\CHANGE{Although single-droplet FESEM images on SiO\textsubscript{2}/Si (Fig.~S10(f)) show a ring-like profile indicative of a coffee-ring effect, the actual coatings on the CMOS chip are formed by overlapping many droplets to create continuous films over ensembles of pixels. The resulting lateral response maps in Fig.~5(b–e) and the averaging over selected regions (Fig.~5(f–g)) demonstrate that, at the scale of the sensing pixels, the coatings are effectively homogeneous and do not exhibit detectable electrical response segregation. Remaining non-uniformities can be corrected for by calibration measurements at the individual pixel level.}


\begin{table}[htbp]
    \centering
    \small
    \caption{Printing recipes for the inks.}
    \label{table:ink-printing}
    \begin{tabular}{|l|c|c|c|c|}
        \hline
        \textbf{Parameter} & \textbf{UV Ink} & \textbf{ZIF-8 Ink} & \textbf{MIL-101(Cr) Ink} & \textbf{MIL-140A Ink} \\
        \hline
        Printhead & DMC 11610 & DMC 11610 & DMC 11610 & DMC 11610 \\
        Voltage Raw B & 40 & 40 & 40 & 100 \\
        Print speed (mm/s) & 50.8 & 50.8 & 50.8 & 50.8 \\
        Ink pressure (mbar) & -1 & -6 & -5 & -5 \\
        Resolution (DPI) & 500 & 500 & 500 & 500 \\
        Native droplet volume (pL) & 10 & 10 & 10 & 10 \\
        Measured droplet volume (pL) & 9 & 8.9 & 9.2 & 9.3 \\
        Nozzles used during printing & 1 & 1 & 1 & 1 \\
        \hline
        \multicolumn{5}{|c|}{\textbf{Pulse Shape Voltages (V)}} \\
        \hline
        High Voltage & 27 & 25 & 20.0 & 26 \\
        Medium Voltage & 9 & 13 & 13.0 & 9 \\
        Low Voltage & 3 & 3 & 1.5 & 3 \\
        \hline
        \multicolumn{5}{|c|}{\textbf{Pulse Shape Time Duration ($\mu$s)}} \\
        \hline
        Up Down & 0.1 & 0.1 & 20.0 & 20.0 \\
        Idle Before & 6 & 0.5 & 3.0 & 0.5 \\
        Fill Ramp & 5 & 0.5 & 3.0 & 3.0 \\
        Time Low & 6 & 1 & 6.0 & 10.0 \\
        Fire Ramp & 2 & 4 & 2.4 & 2.4 \\
        Time High & 3 & 0.5 & 2.0 & 2.0 \\
        End Ramp & 5 & 0.5 & 2.0 & 2.0 \\
        Idle After & 7 & 10 & 5.0 & 5.0 \\
        \hline
    \end{tabular}
\end{table}

\subsection{VOCs Gas Sensing Setup of Pure and Mixed VOCs}
\label{sec:voc-sensing-setup}

\begin{figure}[htbp]
    \centering
    \includegraphics[width=0.7\textwidth]{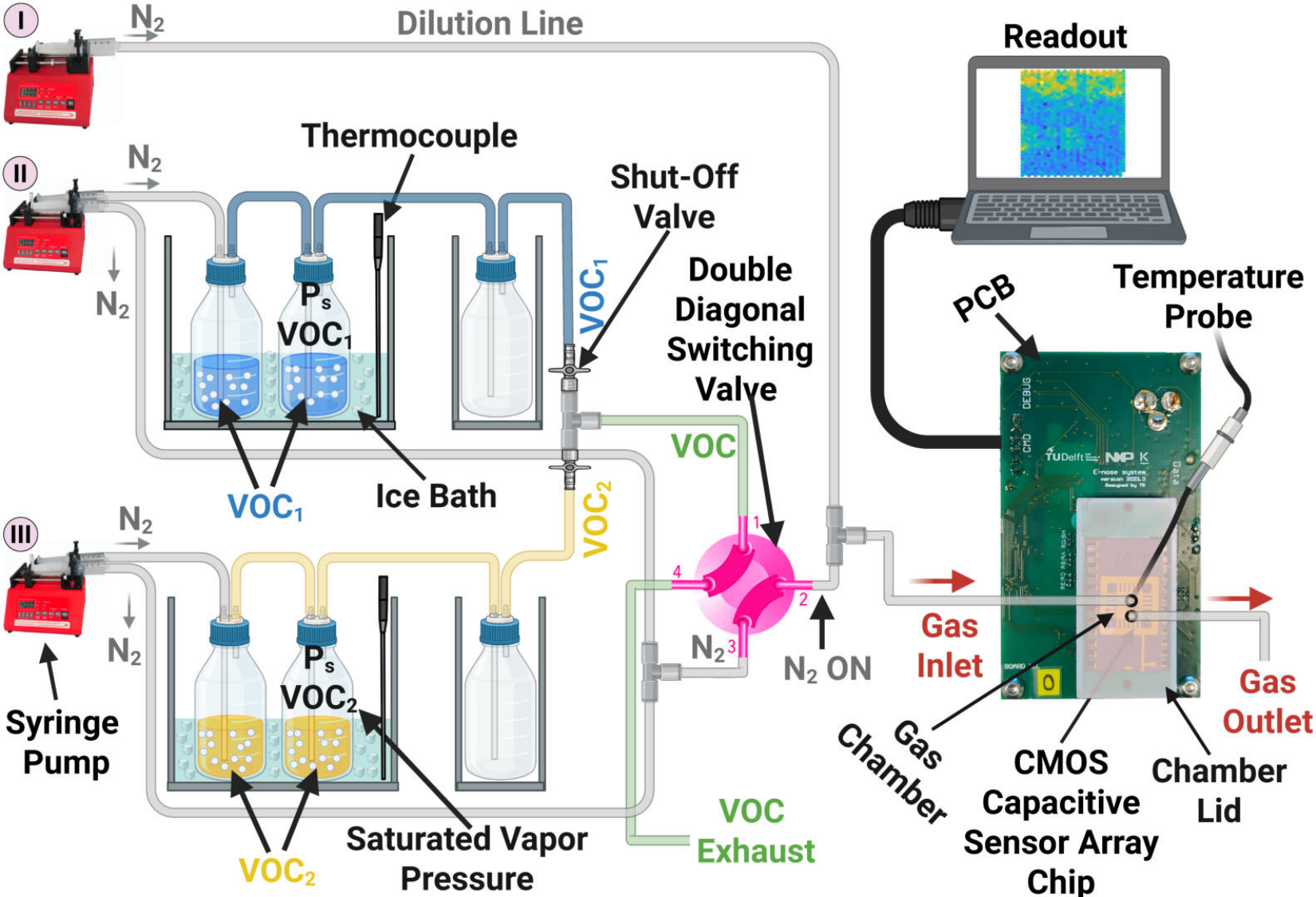}
    \caption{
        \textbf{Pure and mixed VOCs gas sensing setup.} The setup comprises three syringe pumps (I--III) that deliver \(\mathrm{N}_2\) to the dilution, \(\mathrm{VOC}_1\), and \(\mathrm{VOC}_2\) channels, respectively. Each VOC channel includes a shut-off valve to select \(\mathrm{VOC}_1\), \(\mathrm{VOC}_2\), or their mixture. A four-way diagonal valve (shown in pink) switches the flow sent to the sensor between pure \(\mathrm{N}_2\) and \(\mathrm{N}_2\) mixed with VOCs. The schematic depicts the pure \(\mathrm{N}_2\) case. To expose the chip to a \(\mathrm{VOC}_1/\mathrm{VOC}_2\) mixture, rotate the diagonal valve by \(90^{\circ}\) to enable VOC delivery. The experimental setup is shown in Fig.~\ref{fig:lab-setup}.
        }
    \label{fig:gas-sensing-setup}
\end{figure}

In this study, the pixelated capacitive sensor (PCS) CMOS platform detects VOCs via adsorption-induced permittivity changes in inkjet-printed MOF/polymer layers on capacitive pixels. This capacitive transduction is fundamentally different from the O$_2$-driven band-bending mechanism of resistive metal-oxide sensors \cite{zaki2019role}. Consequently, under dry conditions, dry synthetic air (O$_2$ in N$_2$) and dry N$_2$ are expected to yield equivalent PCS responses. We verified this at the outset (dry synthetic air vs. dry N$_2$) and observed indistinguishable signals; therefore, all subsequent measurements used dry N$_2$ as the carrier at constant total flow. 

The CMOS sensor chip operated at a self-heated steady temperature of \(30 \pm 1\,^{\circ}\mathrm{C}\), measured on the package adjacent to the die (thermocouple). The chamber air was \(23 \pm 2\,^{\circ}\mathrm{C}\). 

The VOC gas sensing experiments were conducted using a controlled exposure system, as illustrated in Fig.~\ref{fig:gas-sensing-setup}. The system comprises a VOC bubbler, a dilution stage, and a test chamber where the sensor is positioned for real-time measurements. This setup is designed to test both pure and mixed VOCs under controlled environmental conditions. The switching time remains fixed throughout the experiment to systematically evaluate the ON/OFF behavior of the sensor. Fig.~\ref{fig:lab-setup} shows a photograph of the actual experimental gas sensing setup.

The setup operates by pumping N$_2$ through three individual pumps, numbered I-III, using independent syringe pumps (NE-4000 Two Channel). The setup operates by delivering dry N$_2$ via three independently controlled syringe pumps (NE-4000 Two Channel), designated I–III. The carrier gas is dry N$_2$ at a constant total flow (maintained throughout the experiments).

 Pump II is designated for VOC$_1$ (for pure gas measurements), while pump III is reserved for VOC$_2$ (in cases of mixture experiments). The entire setup utilized standard microfluidic polytetrafluoroethylene (PTFE) tubing (1/16'' OD $\times$ 1/32'' ID). Pump I is used for dilution in both pure and mixing experiments. The flow rate over the CMOS capacitive sensor array chip is kept constant at 2 mL/min throughout the entire experiment.
Note that pumps~II and~III each contain two identical syringes filled with \(\mathrm{N}_2\). Consequently, pump~II and pump~III deliver equal volumetric flow rates, \(\dot{V}_2\) and \(\dot{V}_3\), respectively, which are split into two branches: one branch passes through the VOC bubbler, while the other bypasses the bubbler(s).

To efficiently generate vapor at the saturated pressure (\(P_s\)), we used a two-stage bubbling train (glass bottles) followed by a third, empty knock-out bottle to prevent VOC condensate from reaching the sensor chip. The physical and chemical properties of all VOCs in this study, including their \(P_s\), are summarized in Table~\ref{tab:VOC_properties}. Where all solvents were of high purity: Ethanol (\(\geq 99.5\%\), Thermo Fisher) and 2-Propanol (\(\geq 99.7\%\), Sigma-Aldrich), Methanol (\(\geq 99.9\%\), Honeywell), 1-Butanol (\(\geq 99.4\%\), Sigma-Aldrich), 2-butanone (\(\geq 99.5\%\), Sigma-Aldrich), and toluene (\(\geq 99.8\%\), Merck) were used in the gas-sensing studies.

\begin{table}[htb]
    \centering
    \small
    \renewcommand{\arraystretch}{1.3} 
    \caption{
        Physical and Chemical Properties of VOCs. 
        MW: Molecular weight (g/mol), Ps: Vapor pressure at 1°C (mmHg), 
        DC: Dielectric constant, 
        MD: Molecular diameter (\r{A}), BP: Boiling point (°C), 
        DM: Dipole moment (D), P: Polarity, SI: Solubility interactions.
    }
    \label{tab:VOC_properties}
    \begin{tabular}{l c c c c c c c p{2.5cm}}
        \toprule
        \textbf{VOC} & \textbf{MW} & \textbf{Ps}\cite{yaws2005yaws} & \textbf{DC}\cite{akerlof1932dielectric} & \textbf{MD} & \textbf{BP} & \textbf{DM} & \textbf{P} & \textbf{SI}\cite{grate1991solubility} \\
        \midrule
        \textbf{Ethanol}   & 46.07  & 15.23  & 24.55  & $\sim$4.4  & 78.0  & 1.69 & Polar & \textbf{Dipolarity, Hydrogen Bonding} \\
        \textbf{2-IPA}     & 60.10  & 8.96   & 18.3   & $\sim$4.8  & 82.0  & 1.66 & Polar & \textbf{Dipolarity, Hydrogen Bonding} \\
        \textbf{Methanol}  & 32.04  & 31.51  & 32.7   & $\sim$4.0  & 65.0  & 1.70 & Polar & \textbf{Acidity, Dipolarity} \\
        \textbf{Water}     & 18.02  & 4.89   & 78.3   & $\sim$3.1  & 100.0 & 1.85 & Highly Polar & \textbf{Strong Hydrogen Bonding, Acidity} \\
        \textbf{2-butanone} & 72.11  & 26.70  & 18.5   & $\sim$4.5  & 79.6  & 2.78 & Polar & \textbf{Dipolarity, Weak Hydrogen Bonding} \\
        \textbf{1-Butanol}  & 74.12  & 1.72   & 17.8   & $\sim$5.1  & 117.7 & 1.66 & Polar & \textbf{Dipolarity, Hydrogen Bonding} \\
        \textbf{toluene}   & 92.14  & 7.54   & 2.38   & $\sim$5.8   & 110.6 & 0.37 & Non-Polar & \textbf{Polarizability, Dispersion Forces} \\
        \bottomrule
    \end{tabular}
\end{table}

The experiments were conducted at a constant temperature of 1°C, maintained using an ice bath, and precisely controlled via a thermocouple. By controlling the temperature, we can estimate the saturated vapor pressure of the gas using the Antoine equation (see Sec.~\ref{sec:antoine-equation}), and use this to estimate the VOC concentration in the gas flow. A third bottle after the 2 bubbler is used to bring the gas to room temperature. Before each experimental run, pure dried nitrogen was flushed through the system for one hour to remove any residual gases and prevent humidity accumulation. Simultaneously, the bubblers were activated to ensure the VOCs reached their saturated vapor pressure before entering the measurement setup. We change the 
volume flow rates of the pumps to reach the desired concentrations as explained in equations Eqs.~\ref{eq:general-concentration} and \ref{eq:antoine}.

During the measurement process, a four-way diagonal valve (IDEX 4-Port Switching Valve) was used to switch between exposing the sensor to VOC vapors and pure N$_2$.
The valve and pump settings are configured such that we ensure the gas flow rate over the sensor is constant during the experiment, with only the composition of the gas being changed. This reduces flow-rate-dependent effects on the sensor response. Multiple microfluidic Y connectors made of PEEK were used to connect the different lines. 



During the gas sensing experiments, the gas concentrations and of both VOC1 and VOC2 are varied by controlling the flow rates $\dot{V}_2$ and $\dot{V}_3$ using syringe pumps II and III, while the flow rate $\dot{V}_1$ of pump I is adjusted to maintain the total gas rate over the sensor constant at $\dot{V}_{\text{total}}= 4 \text{ mL/min}$ according to this equation:

\begin{equation}
    \dot{V}_{\text{total}} = \dot{V}_1 + \dot{V}_2 + \dot{V}_3 = 4 \text{ mL/min}.
\end{equation}

To calculate the concentration of each of the VOCs on the sensor surface, we first need to calculate its concentration after it leaves the bubbler. For this purpose, we use the Antoine equation to determine the saturated vapor pressure $P_s$ of the VOC at the temperature of the ice, and then use the following equation to determine the VOC concentration that reaches the sensor:







\begin{equation}
    C_{\mathrm{VOC1}}= \frac{P_{s,1}}{P} \times \frac{\dot{V}_2}{\dot{V}_\mathrm{total} } \times 10^6~ \mathrm{ppm}
    \label{eq:general-concentration}
\end{equation}
Where $C_{\mathrm{VOC1}}$ is the VOC concentration, $P_{\text{s,1}}$ is the saturated vapor pressure of VOC1 at 1~$^\circ$C as determined from the Antoine equation (see Sec.~\ref{sec:antoine-equation}), and $P$ is the atmospheric pressure. The first factor in Eq.~\ref{eq:general-concentration} accounts for the concentration of the VOC in the bubbler, while the second factor accounts for the dilution by mixing with the nitrogen flow. For VOC2 the equation becomes $C_{\mathrm{VOC2}}= \frac{P_{s,2}}{P} \times \frac{\dot{V}_3}{\dot{V}_\mathrm{total} } \times 10^6~ \mathrm{ppm}$.

\section{Results}
\subsection{Working Principle and E-nose chip electronics}

\begin{figure}[htbp]
    \centering
    \includegraphics[width=0.9\textwidth]{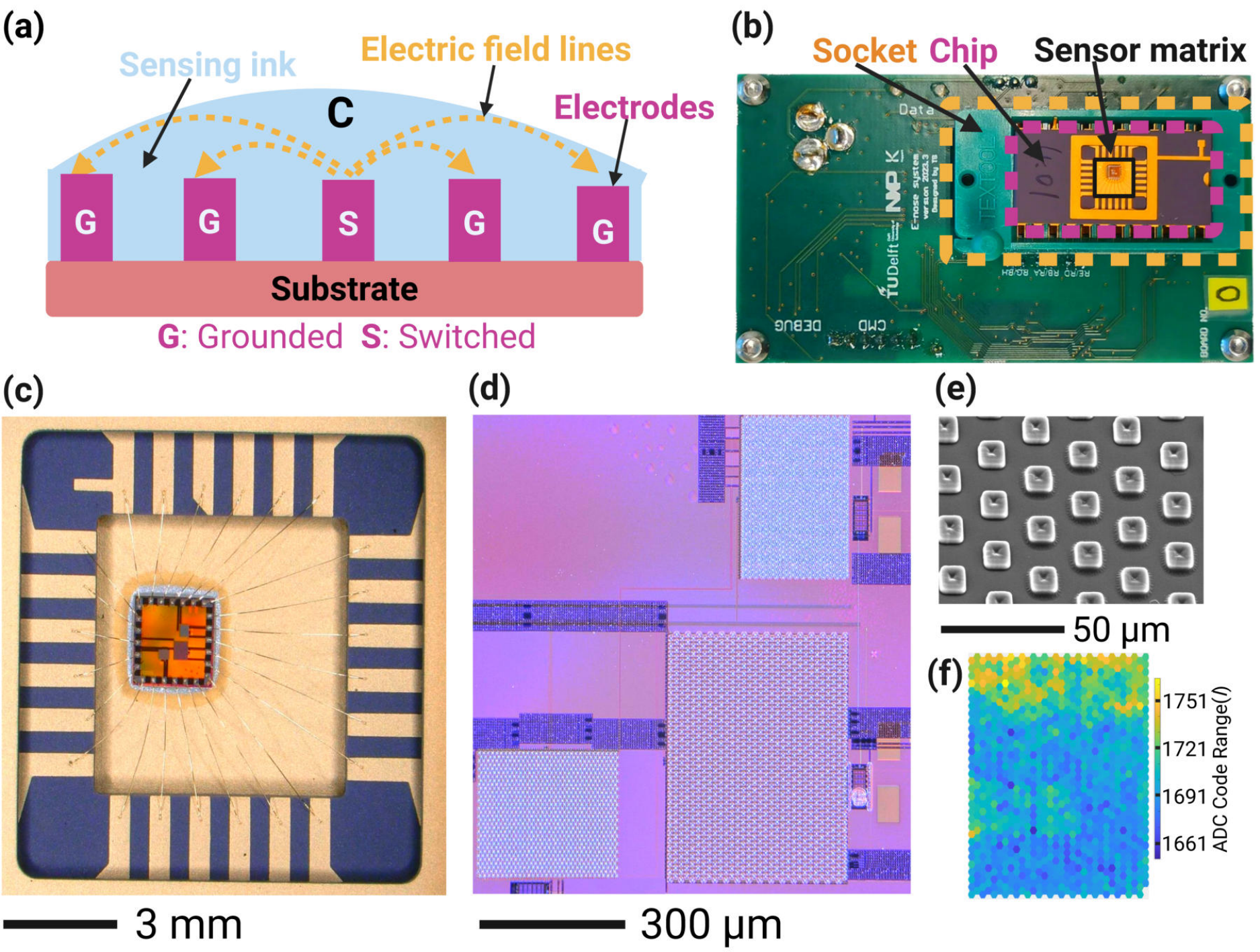}
    \caption{
        \textbf{Working principle and CMOS pixelated sensing chip.}  
        \textbf{(a)} Working principle of the capacitive gas sensor. The capacitance changes when exposed to VOCs due to the effect of the gas on the dielectric constant of the sensing ink. A more detailed CMOS device cross-section is shown in Fig.~\ref{fig:Xsection}.
        \textbf{(b)} Packaged CMOS pixelated sensing chip mounted on readout PCB
        \textbf{(c)} Close-up optical image of the CMOS pixelated chip in a ceramic package.  
        \textbf{(d)} Microelectrode arrays of the pixelated capacitive sensing chip. The dimensions of the microelectrodes in the three sensor matrices are shown in Fig.~\ref{fig:chip-microelectrodes}.  
        \textbf{(e)} SEM top-view image showing the pixelated microelectrode structures.  
        \textbf{(f)} Display of the capacitance distribution over all 1024 pixels, with corresponding ADC values indicated by the colorscale. 
    }
    \label{fig:working-principle-chip}
\end{figure}

The capacitive E-nose operates by detecting capacitance variations between its electrodes that are caused by changes in the dielectric properties of the sensing ink upon exposure to VOCs. The working principle and CMOS pixelated chip design are illustrated in Fig.~\ref{fig:working-principle-chip}, showcasing both the functional mechanism and the hardware integration.

\begin{figure}[htbp]
    \centering
    \includegraphics[width=0.9\textwidth]{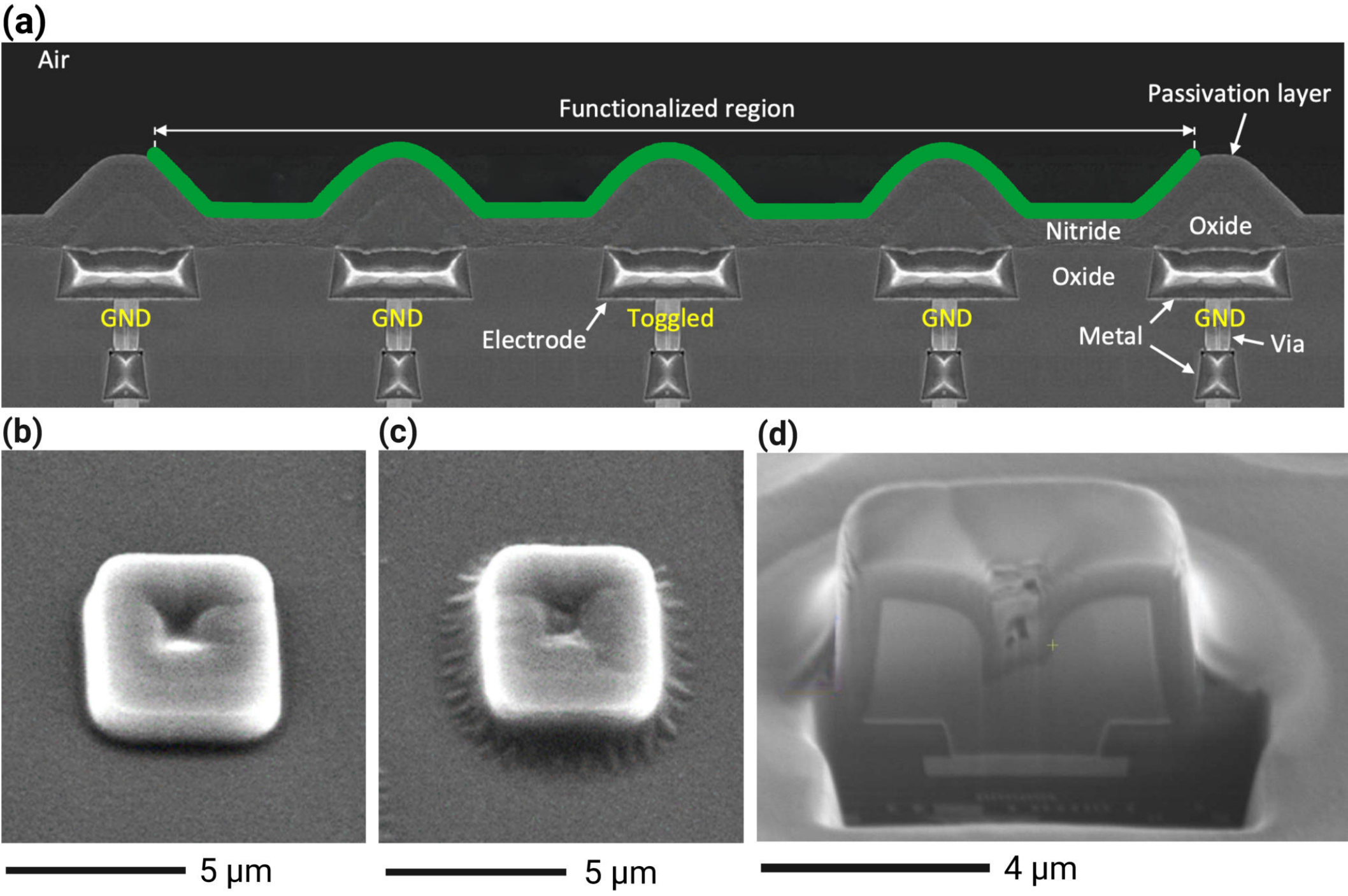}
    \caption{\textbf{Artist impression, based on SEM images, of cross-section of CMOS sense electrodes by sensing ink.}
    (a) The electrodes, made in the top-metal layer of the CMOS process, are connected vertically by vias and intermediate metal layers to pairs of switching transistors at the bottom (not visible). The middle electrode is selected by toggling (switching) its potential between $0$ and $0.9\,\text{V}$ at high frequency. This makes it the active plate of the selected sense capacitor. Electric field lines emerging from it pass through the green sensing ink and terminate mainly on grounded adjacent electrodes that collectively constitute the counter electrode plate of the selected sense capacitor. Note that Fig. (a) is an artist's impression that has been constructed by cut-and-pasting SEM cross-section images from various parts of the chip. (b) SEM image of a reference bare electrode. (c) SEM image of a functionalized electrode with a UV ink. Note that one ink droplet covers several electrodes. 
    (d) A focused-ion-beam (FIB) SEM cross-section image of a single electrode pixel. Thinner CMOS backend metal layers that connect the pixel to the transistors are visible below the thick top electrode. An ink meniscus is adhering to the edges on both sides of the pixel.}
    \label{fig:Xsection}
\end{figure}

The operation mechanism is illustrated in Fig.~\ref{fig:working-principle-chip}(a). Under reference conditions, where only N$_2$ gas surrounds the sensor, the sensor detects a reference capacitance value ($C_0$). However, when exposed to VOCs, interactions of the gas with the sensing ink lead to a change in the effective dielectric constant \cite{fernandez2021printed}, thereby changing the capacitance to ($C_0 + \Delta C$). 
Microscopically, this capacitance change can originate from changes in the charge, conductivity, and dielectric constant distributions in the ink that change its effective dielectric constant when the gas molecules diffuse through the material or adsorb on the porous surface of the functional layer. By engineering functional layers whose effective capacitance is affected more by certain gases than by others, the sensor's selectivity can be enhanced. The capacitance changes are detected by switching the voltage on the switch (S) electrode between 0 and 0.9 V at a frequency of 40 MHz. This causes a capacitive current proportional to the capacitance between the S and G (grounded) electrodes (See Fig. 1a), which is digitized by an analog-to-digital converter (ADC) in the CMOS chip.

In this work, we describe the CMOS pixelated sensing chip and readout electronics that we designed, fabricated used for gas sensing, utilizing the described capacitive operation principle. Photographs and micrographs of the capacitive pixelated CMOS E-nose are shown at different magnifications in Figures~\ref{fig:working-principle-chip}(b-e) and in Fig.~\ref{fig:Xsection}.
Fig.~\ref{fig:working-principle-chip}(b) presents the printed circuit board (PCB), with a socket for mounting the packaged CMOS pixelated sensing chip, providing a USB connection between the sensor array and a computer.
Fig.~\ref{fig:working-principle-chip}(c) shows an optical image of the CMOS chip that is mounted in a ceramic package and electrically connected by bondwires.

\CHANGE{The pixelated capacitive sensor array that resembles our work most is the recently reported MOF-integrated e-nose that realizes high-density pixel arrays with a pixel dimension of 45$\times$45 $\mu$m$^2$ (electrode spacing $\sim$2 $\mu$m), operating at 156.25 kHz \cite{saif202520}. The platform features 4,096 pixels with integrated temperature sensors and heaters for recovery. In this work we present a platform that features 3 pixelated matrices, each having 1,024 capacitive pixels, operating at 40 MHz, enabling detecting capacitance changes less than 1 aF. The smallest pixel dimensions are 4$\times$4 $\mu$m$^2$ with a pixel spacing of 10$\mu$m. Furthermore, all sensing layers are deposited by solution-based inkjet printing directly onto the passivated CMOS surface, providing a wet-process-compatible and purely additive route to multi-material functionalization. The same printing workflow can be extended to larger arrays and a broader range of inks without
additional lithography or etching steps, which is advantageous for low-cost, scalable deployment of high-diversity e-nose systems.}

Fig.~\ref{fig:working-principle-chip}(d) highlights the three pixelated microelectrode sensing matrices on the chip, which serve as the detection interface where the sensing functional layer interacts with VOCs. This figure shows three distinct sensor matrices, each designed with different electrode geometries to support various sensing configurations. Sensor Matrix~1  (bottom right) consists of square electrodes with dimensions of $5 \times 5~\mu\mathrm{m}$ and an interelectrode spacing of $15~\mu\mathrm{m}$. Sensor Matrix~2 (top right) features electrodes with dimensions of $4 \times 4~\mu\mathrm{m}$ and an interelectrode spacing of $10~\mu\mathrm{m}$, while Sensor Matrix~3 (bottom left) has electrodes of $5 \times 5~\mu\mathrm{m}$ with an interelectrode spacing of $10~\mu\mathrm{m}$. These different sensor matrices were designed to investigate and optimize the electrode design. The experiments in this work were performed using Sensor Matrix~1. Details on the electronic design and CMOS fabrication process can be found in Refs.~\citenum{widdershoven2024pixelated, laborde2015real}.

Fig.~\ref{fig:working-principle-chip}(e) displays a top-view SEM image of the microelectrodes in Sensor Matrix~1, and Fig.~\ref{fig:Xsection} an image of their cross-section, illustrating the structural topography of the sensing electrodes. The undulating surface of the CMOS chip, characterized by valleys around the electrodes, is a direct result of the CMOS fabrication process, where silicon-oxide and silicon-nitride passivation layers protrude upward at the location of the electrodes. The capillary forces in the valleys between these protrusions help localize the printed ink in the areas between the microelectrodes, as will be explained in the next section.



Fig.~\ref{fig:working-principle-chip}(f) demonstrates the pixelated sensor readout, showing a spatial map of the measured response by each of the pixelated electrodes under exposure to VOC. The color scale-bar indicates the ADC capacitance readout value under VOC exposure. The ink-coated areas exhibit a stronger capacitance change (yellow pixels) compared to the non-coated areas (dark blue pixels).

To electronically read-out the pixelated sensor, the E-nose chip contains analog-to-digital converters (ADC) that convert the capacitive signals to digital ADC values that are sent by the chip via the PCB and USB connection to the computer. For this purpose, the capacitive current is amplified by a factor of 100 by a current mirror, after which the voltage it generates across a 68 k$\Omega$ resistor is digitized by a 12-bit (4096 values) ADC. With respect to a calibrated reference ADC value $\text{ADC}_{\text{ref}}$, the relationship between the digital ADC readout and the measured capacitance \( C_{\text{measured}} \) of the sensor is given by:

\begin{equation}
    C_{\text{measured}} = \frac{1}{100 \times \SI{0.9}{\volt} \times \SI{40}{\mega\hertz} \times \SI{68}{\kilo\ohm}} \times
    \left[ \frac{(\text{ADC}_{\text{output}} - \text{ADC}_{\text{ref}}) \times \SI{3.3}{\volt}}{4096} \right]
    \label{eq:measured_capacitance}
\end{equation}

According to this equation, a change of 1 in the value of $\text{ADC}_{\text{output}}$ corresponds to a capacitance change of 3.3 aF of the sensor, and the 12-bit capacitance detection range is 4096$\times$3.3 aF=13.5 fF. Note that by averaging, even smaller capacitance changes can be achieved, as shown in section~\ref{sec:capacitance-derivation}. A dedicated MATLAB code has been developed to analyze pixelated responses, as shown in Fig.~\ref{fig:data-analysis}, and to generate figures such as Fig.~\ref{fig:working-principle-chip}(f). The software enables continuous analysis of the average capacitance change over user-selected sets of pixels, e.g., to separately analyze and compare the effect of gases on functionalized and non-functionalized electrode areas in different parts of the pixelated array.

\CHANGE{The effective thermal noise is reduced by averaging the responses of multiple pixels covered by the same functional layer and by applying a moving average over time to the resulting traces. No further digital filtering is used, thereby preserving all relevant dynamic features of the sensor response. From repeated measurements in nitrogen, the standard deviation of the residual noise in Figure~7
corresponds to an input-referred capacitance fluctuation of approximately 1~aF per sensor element, which is already very small compared to the baseline capacitance.}

\subsection{Sensor functionalization by inkjet printing}
\begin{figure}[htbp]
    \centering
    \includegraphics[width=0.9\textwidth]{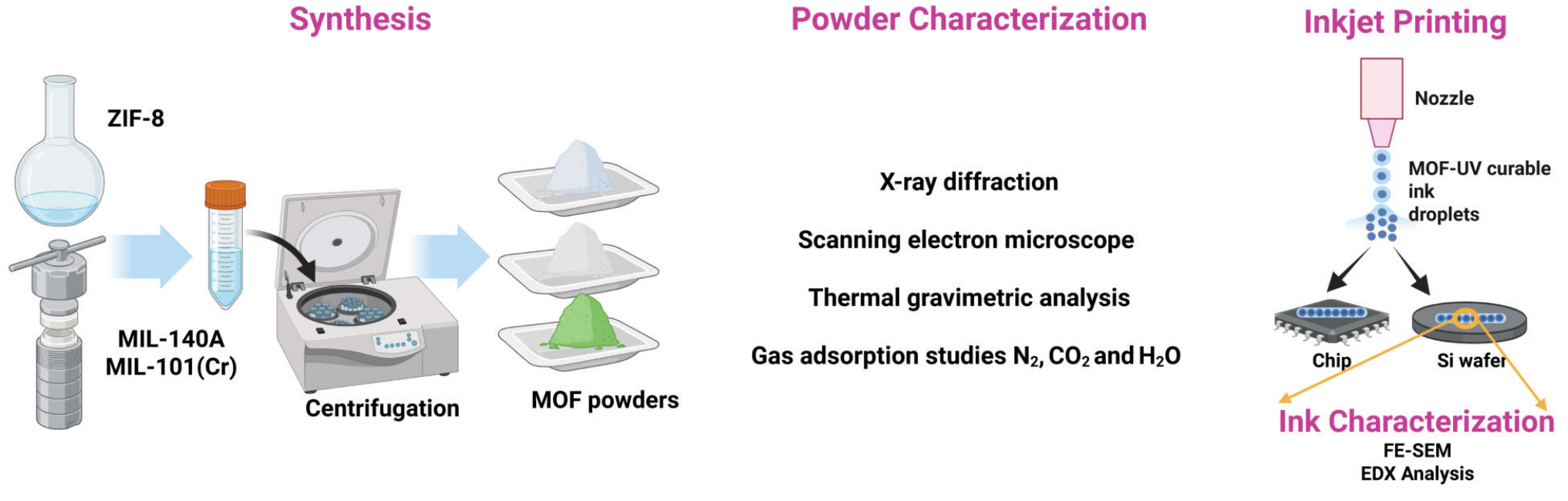}
    \caption{
        \textbf{Graphical summary of the material synthesis, deposition, and characterisation}. Various MOF powders, which are selected for their gas sensing potential, were synthesized, characterized, and mixed with ethanol/water solvent to make them printable ink. Later, it was mixed with UV-light curable ink to improve the stiction of the MOFs to the sensor array. Droplets of the mixed inks containing the MOF powder are inkjet printed on the sensor chip, UV cured, and characterized. The details of these steps, and the improvement of the ink formulation, can be found in Sec.~\ref{sec:experimental}.
    }
    \label{fig:materials-summary}
\end{figure}

The pixelated capacitive sensor array is functionalized via direct inkjet printing of pre-synthesized MOF \cite{najafabadi2025advancements}. The functionalization workflow is summarized in Fig.~\ref{fig:materials-summary}. It illustrates Metal-Organic Framework (MOF) synthesis, characterization, ink formulation, inkjet printing, and printed ink analysis. 
 The details of the material preparation and characterization are described in Sections~\ref{sec:experimental}, \ref{sec:mof-nanoparticles-characterization}, and \ref{sec:surface-structural-characterization}. The synthesized MOFs, namely ZIF-8 (\textit{zeolitic imidazolate framework-8}), MIL-101(Cr) (\textit{Materials of Institut Lavoisier-101, chromium; Cr(III) terephthalate}), and MIL-140A (\textit{Materials of Institut Lavoisier-140A; Zr(IV) terephthalate}) were structurally analyzed using X-ray diffraction (XRD), scanning electron microscopy (SEM), and Thermogravimetric analysis (TGA) before being formulated into Ultraviolet (UV)-curable inks for printing onto Silicon wafer and chip. The printed structures were characterized by field emission scanning electron microscopy (FE-SEM) and energy-dispersive X-ray (EDX) analysis.
To evaluate sensor selectivity, we tested a panel of VOCs and additionally measured hydrophobicity as explained in the next sections. 
 
The structural and adsorption analyses of the MOFs ZIF-8, MIL-101(Cr), and MIL-140A are discussed in detail in Section \ref{sec:mof-nanoparticles-characterization}. Further discussion focusing on ZIF-8 and the UV-curable inks is provided in more detail here.
To functionalize the chip, ZIF-8 ink was pre-mixed with UV ink at a 2:98\% mixing ratio. Initially, inkjet printing parameters were optimised for droplet shape, distribution, and diameter by printing on SiO$_2$/Si wafers. The optical microscopy images of both inks are shown in Fig.~\ref{fig:inkjet-characterization}, revealing well-patterned circular droplets with average diameters of $29.9 \pm 0.2\,\si{\micro\meter}$ for the UV ink 
and $64.7 \pm 0.9\,\si{\micro\meter}$ for the ZIF-8 ink 


To further assess printing uniformity, additional patterned arrays were inkjet-printed on SiO$_2$/Si wafers for both inks, as shown in Fig.~\ref{fig:inkjet-patterns}. These patterned matrices demonstrate a uniform droplet distribution, highlighting the inkjet printing process's repeatability, uniformity, and precision.
Surface characterization of the individual droplets containing ZIF-8 nanoparticles was carried out using FE-SEM, as shown in Fig.~\ref{fig:FESEM-ZIF8}. The FE-SEM images confirm that the ZIF-8 nanoparticles retain their expected morphology, consistent with their structure before printing, as depicted in Fig.~\ref{fig:ZIF8-structure}. 
EDX analysis was conducted on the total EDX intensity images of the inkjet-printed UV and ZIF-8 inks. Elemental mapping results of a selected area for UV and ZIF-8 inks are shown in Fig.~\ref{fig:EDX-analysis} (c and d), confirming the successful printing and uniform elemental distribution within the deposited inks.

EDX analysis was conducted to investigate the total EDX intensity image of the inkjet-printed UV and ZIF-8 inks. Selected area analysis and the corresponding spectrum for the UV ink (Fig.~\ref{fig:EDX-analysis} (a and c)) confirm the presence of carbon (C), nitrogen (N), and oxygen (O), as expected for the organic matrix. For ZIF-8, the grayscale elemental mapping in Fig.~\ref{fig:EDX-analysis} (b and d) highlights the distribution of zinc (Zn), O, and C across the printed area at high magnification. Zn appears to be well-distributed throughout the scanned region.

\begin{figure}[htbp]
    \centering
    \includegraphics[width=0.7\textwidth]{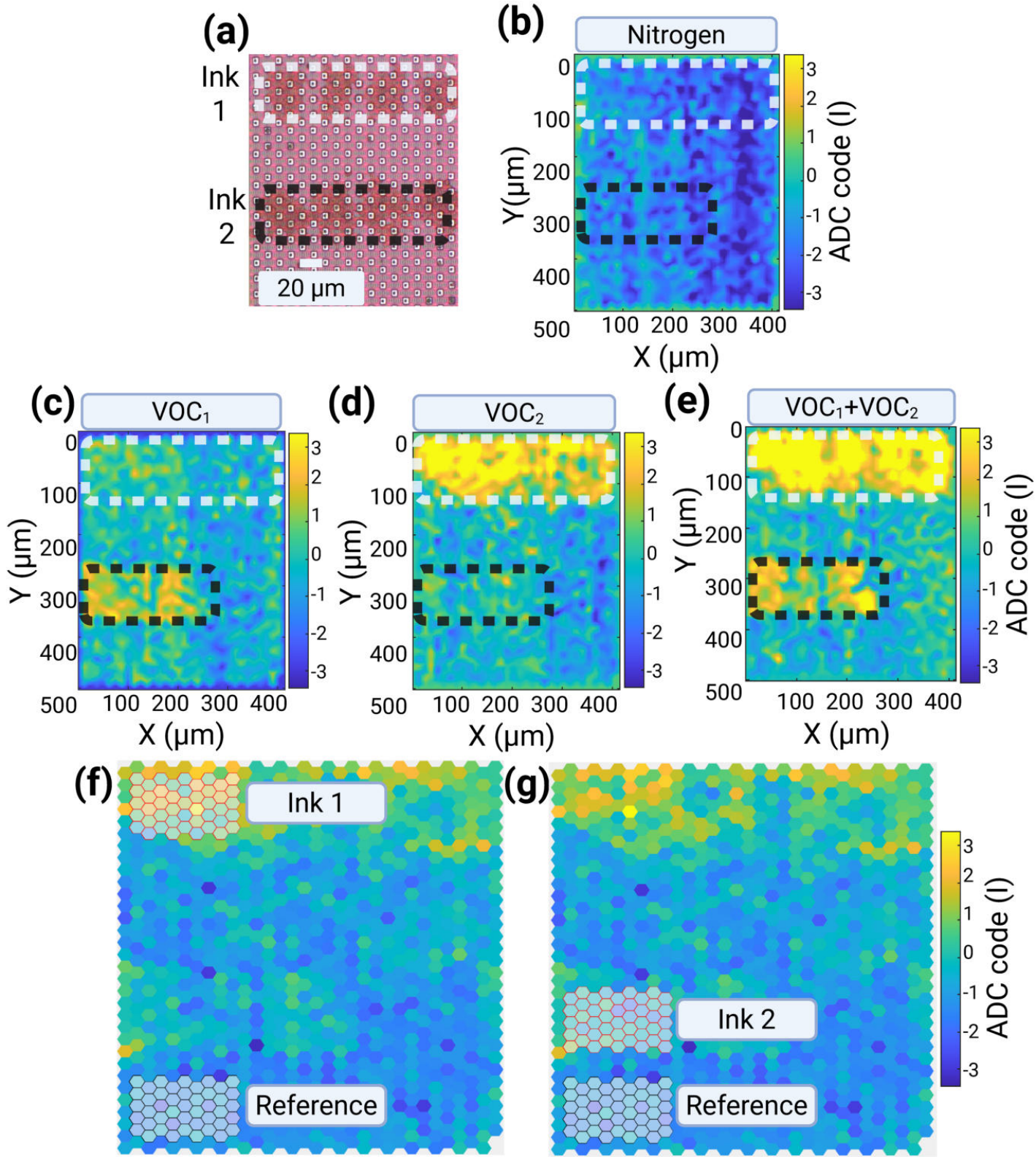}
    \caption{
        \textbf{Sensor response of the pixelated sensor matrix to VOCs after functionalization by inkjet printing.}  
        (a) Optical microscopy image of inkjet-printed ZIF-8 (ink~1) and UV-curable ink (ink~2) single droplets spaced by 15~\(\mu\text{m}\) on part of the CMOS microelectrode array. Details of the microelectrode-array structures for the different sensor matrices are explained in Fig.~\ref{fig:chip-microelectrodes}.
        The image shows individual droplets after UV light curing, followed by multiple droplet depositions, leading to their merging as in (b-e). 
        (b–e) Lateral distribution of the capacitance change $\Delta \mathrm{ADC_{output}}$ under gas conditions. The scale bar represents the change in the ADC code (I), where 1 ADC code represents a capacitance change of 3.3 aF as expressed in Equation 1. 
        The dashed lines indicate different ink regions: ZIF-8 ink (white) and UV ink (black). This sensor chip is different from that shown in (a).
        (b) Response under N$_2$ (reference).  
        (c) Exposure to toluene.   
        (d) Exposure to 2-butanone.   
        (e) Mixed toluene and 2-butanone exposure.
        (f-g) Selected regions over which ADC values are averaged for determining $\Delta \text{ADC}_{\text{output,$t_0$}}$ for ink 1 and ink 2 functionalised areas and $\Delta \text{ADC}_{\text{output,empty}}$ in the reference region.
    }
    \label{fig:inkjet-print}
\end{figure}

After optimizing the printing conditions on Si wafers, the inkjet printing process was applied to CMOS microelectrode chips. Fig.~\ref{fig:inkjet-print}(a) shows 4 droplets of UV ink and 4 droplets of ZIF-8 ink deposited on Sensor Matrix 1 by inkjet printing; the droplets are observed to cover roughly circular regions on the chip having a diameter of around 50 $\mu$m. The layer thickness and surface coverage are not always optimal, as is e.g. visible in the UV ink droplet on the bottom-right of Fig.~\ref{fig:inkjet-print}(a). Therefore, to increase ink layer height and chip coverage uniformity, we deposit multiple ink droplets at the same position to improve sensing signals.

\subsection{Measurement method}

 To illustrate the basic operation principle of the pixelated capacitive sensor (PCS) array, Figs.~\ref{fig:inkjet-print}(b–e), illustrate the spatial distribution of the capacitance change $\Delta C$ under various gas exposure conditions. The colorscale indicates $\Delta \mathrm{ADC}_{\text{output}}
= \mathrm{ADC}_{\text{output}}\, (t)
- \mathrm{ADC}_{\text{output}}\, (t_0)$, the difference between $\mathrm{ADC}_{\text{output}}\, (t)$ and the reference at $t_0$, the difference of the value of $\mathrm{ADC_{output}} (t)$ compared to its reference value in nitrogen gas at atmospheric pressure at the start of the measurement sequence at time $t_0$.  Fig.~\ref{fig:inkjet-print}(b) shows the sensor response in a nitrogen environment, with the observed spatial variations giving a qualitative impression of the sensor stability. Figures~\ref{fig:inkjet-print}(c-d) highlight the localized sensor responses of UV ink and ZIF-8 ink when exposed to toluene (methyl benzene, $C_6H_5CH_3$) and 2-butanone (methyl ethyl ketone, $CH_3C(O)CH_2CH_3)$, respectively. The UV ink shows the strongest response to toluene, while the ZIF8 ink shows the strongest response to 2-butanone.  When both VOCs are present together, as shown in Fig.~\ref{fig:inkjet-print}(e), both inks exhibit a response, confirming the sensor's ability to detect VOCs individually and in a mixture. 
This demonstrates that our pixelated CMOS microarray is a multi-sensing platform for detecting multiple VOCs by integrating different ink materials. Each ink selectively interacts with specific VOCs, enabling continuous detection of multiple gases with the same sensor matrix.

\CHANGE{In the final transducer configuration shown in Fig.~5(b–g), only two functional layers are therefore used on the CMOS array: a ZIF-8 ink region and a region coated with
the neat UV-curable resin, which serves as the polymer reference. MIL-101(Cr) and MIL-140A were employed in the initial humidity-screening experiments (Fig.~6(a–b)) but
were not used in the final VOC calibration and mixture measurements because their strong water uptake leads to large humidity cross-sensitivity outside the scope of the present study.}

To evaluate the response of only specific pixels that are located beneath the ink's functionalized layer, software allows users to manually select the relevant pixels and is programmed to average the value of $\Delta \mathrm{ADC_{output,f0}}$ of the selected pixels Fig.~\ref{fig:inkjet-print}(f-g), which is then plotted in the graph of the sensor response. The same software is also used to select and average the ADC value over an empty, non-functionalized region on the chip to get $\Delta \mathrm{ADC_{output,empty}}$. Then the average and background subtracted response of the functionalized region is calculated using $\Delta \mathrm{ADC_{output,f}}=\Delta \mathrm{ADC_{output,f0}}-\Delta \mathrm{ADC_{output,empty}}$. This averaging procedure over a large number of the (in total) 1024 sensing pixels significantly reduces noise.

\subsection{Demonstration of gas sensing in pure gases and gas mixtures}

A dedicated setup was used to control individual gas concentrations of the two VOCs in a nitrogen gas background flow. Details of this setup and the experimental procedure can be found in the experimental section (Fig.~\ref{fig:gas-sensing-setup}) and section~\ref{sec:voc-sensing-setup}.

This setup measures the humidity responses of various MOFs and evaluates VOC selectivity; details are provided in the following sections.

\subsubsection{Humidity response}
\label{sec:humidityresponse}

\begin{figure}[htbp]
    \centering
    \includegraphics[width=0.7\textwidth]{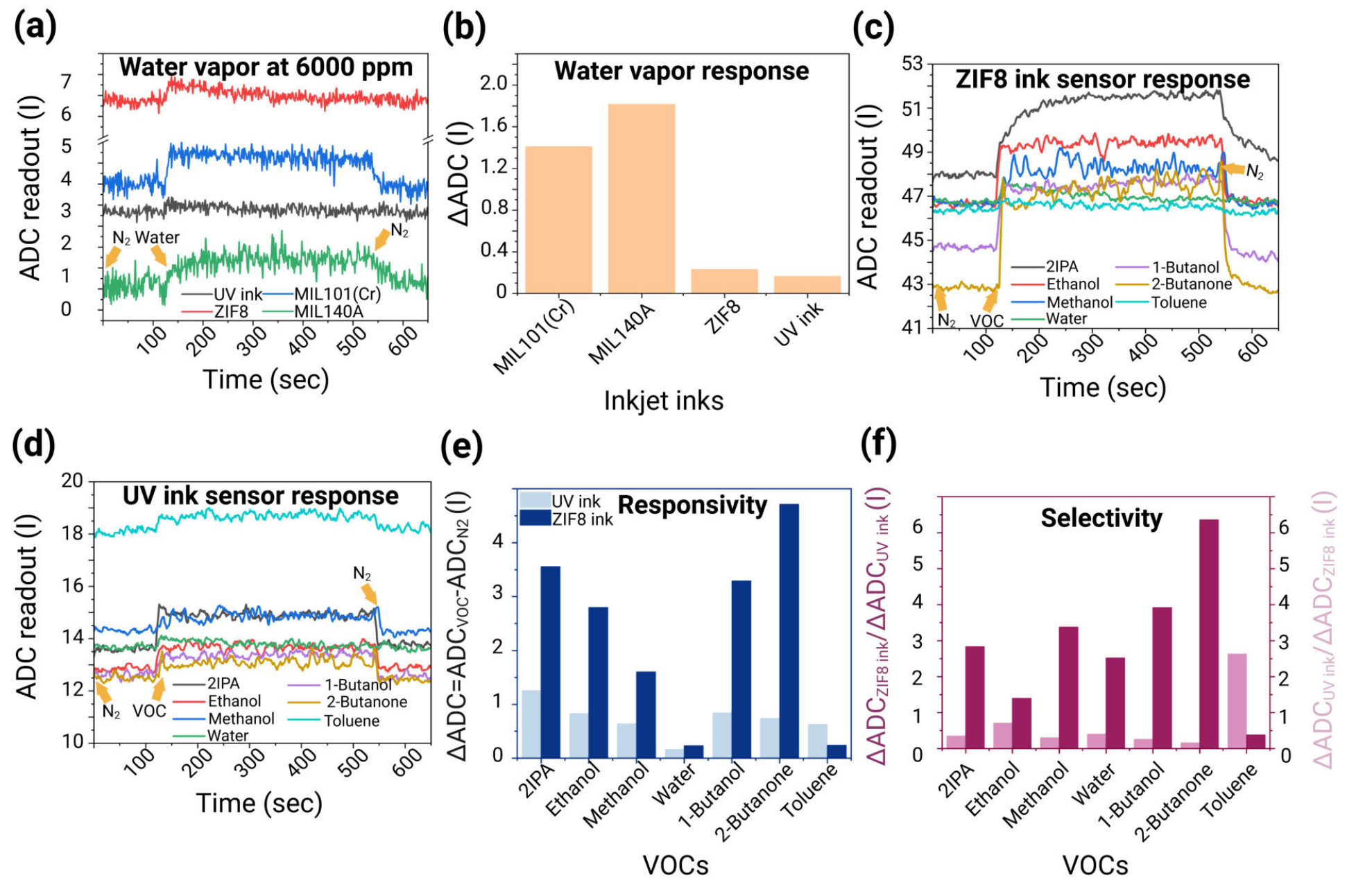}
    \caption{
        \textbf{Water vapor and VOC selectivity analysis of different MOFs.}  
        (a) Dynamic ADC response of MIL-101(Cr), MIL-140A, ZIF-8, and UV ink under exposure to 6000\,ppm water vapor. Arrow lines indicate the switching between the N$_2$ baseline and water vapor exposure. To reduce electronic noise, signals are averaged over a large number of pixels, which affects noise but not the actual sensor response.  
        (b) Quantified water vapor response of inkjet printed samples calculated as the difference in ADC signal between steady-state exposure to water vapor and the baseline under dry N$_2$. Higher values indicate higher water adsorption, reflecting the relative hydrophilicity of the sensing materials.  
        (c) Dynamic sensor responses of UV ink towards different VOCs.  
        (d) Dynamic sensor responses of ZIF-8 ink towards different VOCs.  
        (e) sensor response of UV ink and ZIF-8 ink upon VOC exposure.  Bars show $\Delta\mathrm{ADC}= \mathrm{ADC}_{\text{VOC}}-\mathrm{ADC}_{\mathrm{N_2}}$ (I); light blue corresponds to the UV ink and dark blue to the ZIF-8 ink.  Higher $\Delta\mathrm{ADC}$ values indicate stronger responsivity toward the analyte. \CHANGE{(f) Selectivity of the two inks toward individual VOCs, defined as the ratio of their responses. Dark bars (left axis) plot the ratio $\Delta\mathrm{ADC}_{\text{ZIF-8 ink}}/\Delta\mathrm{ADC}_{\text{UV ink}}$ and identify analytes for which the ZIF-8 ink is more responsive. Light bars (right axis) show the reciprocal ratio $\Delta\mathrm{ADC}_{\text{UV ink}}/\Delta\mathrm{ADC}_{\text{ZIF-8 ink}}$ and identify analytes for which the UV ink is more responsive.}
        }
    \label{fig:hydrophobicity-test}
\end{figure}

For many applications, such as greenhouses and animal farms, it is important that the  E-nose signal is independent of the air humidity level. For this purpose, it is important to verify the response of the sensor to humidity for different functionalization inks. It is expected that hydrophobic sensing materials can reduce the effect of humidity variations on the  E-nose reading.



To check the response to humidity, the sensor was exposed to nitrogen gas that was humidified with 6000 ppm of water vapor. The test was performed for MIL-101(Cr), MIL-140A, ZIF-8, and UV inks. The real-time response curve in Fig.~\ref{fig:hydrophobicity-test}(a) shows the ADC readout of MIL-101(Cr), MIL-140A, ZIF-8, and UV ink. Upon water vapor exposure, MIL-101(Cr) and MIL-140A exhibit a significant increase in response, indicative of a high water adsorption capacity. 

In contrast, ZIF-8 and the UV ink exhibit only minor capacitance shifts, confirming a more hydrophobic nature and lower humidity cross-sensitivity.  This trend matches the water-sorption isotherms of the MOFs: ZIF-8 takes up virtually no H$_2$O until $\sim$80\% RH (Fig.~\ref{fig:ZIF8-adsorption}c), ZIF-8 takes up virtually no H\(_2\)O until \(\sim 80\%\) RH (Fig.~S2(c)), in agreement with previous computational and adsorption studies.\cite{ortiz2014makes, ghosh2014modeling, Coudert2017_JPCC, Hu2012_POC}. Whereas MIL-140A and MIL-101(Cr) adsorb progressively across the RH range (Figs.~\ref{fig:MIL140-adsorption}c and \ref{fig:MIL101-adsorption}c).  

\CHANGE{In these tests, the water concentration of 6{,}000~ppm corresponds to \(\sim 20\%\) relative humidity at 23~\(^\circ\)C, i.e., a typical lower end of indoor humidity. The small capacitance changes observed for ZIF-8 and the UV ink at this level are consistent with the independent powder water adsorption isotherms: ZIF-8 remains essentially dry up to \(\sim 80\%\) RH (Fig.~S2(c)), whereas MIL-101(Cr) and MIL-140A show much higher water uptake (Figs.~S8(c) and S5(c)) and were therefore not used in the final VOC calibration and mixture experiments.}

\CHANGE{The low water response of the UV ink can be rationalized by its composition and cross-linked structure. The cured film is a dense polyester/urethane network formed from multifunctional acrylate monomers and oligomers (Sec.~2.2.2), with a very high aliphatic carbon content (97.5~wt\% C and only 1.6~wt\% O, see Table~S3 and Fig.~S12). This implies a predominantly non-polar environment with only a sparse distribution of polar ester carbonyl groups and hydrogen-bonding sites. The resulting hydrophobic matrix suppresses strong water sorption but still allows appreciable uptake of apolar VOCs such as toluene, as evidenced by the stronger capacitance response of the UV-ink layer to toluene compared to 2-butanone in Fig.~7(g).}

Fig.~\ref{fig:hydrophobicity-test}(a) displays the full time-resolved response, while Fig.~\ref{fig:hydrophobicity-test}(b) shows a histogram of the averaged $\Delta$ADC between 6000~ppm water vapor and the dry N$_2$ baseline.  


These observations can be accounted for by the large hydrophilic cages of MIL-101(Cr) and the 1-D channels of MIL-140A that host open metal sites and hydroxylated surfaces, promoting strong water uptake.  MIL-101(Cr) offers Cr$^{3+}$ coordinatively unsaturated sites, whereas MIL-140A presents Zr–OH groups; both act as potent hydrogen-bond donors/acceptors (see SI Section~S4 for details). 

These results confirm that ZIF-8 and UV ink are advantageous for gas sensing in humid environments because of their lower response to water vapor. Given their reduced water-vapor response, ZIF-8 and the UV ink were chosen for the subsequent gas-sensing measurements.

\subsubsection{VOC response}
\label{sec:voc-response}

E-nose responses to several VOCs were measured to assess the selectivity of the ZIF-8 and UV-based functional layers. \CHANGE{In addition to the MOF-based inks, the neat UV-curable resin (without MOF nanoparticles) was also printed on the CMOS chip and used as a polymer layer. This UV ink was subjected to the same humidity and VOC exposure protocols as the ZIF-8 ink, enabling direct comparison of the polymer matrix response with that of the MOF-loaded composite. No additional polymer matrices were investigated in this study.}
Figures~\ref{fig:hydrophobicity-test}(c-f) present the dynamic responses of ZIF-8 ink towards different VOCs, including 2-propanol (2IPA), ethanol, methanol, water, 1-butanol, 2-butanone, and toluene. The ADC readout is plotted against time, illustrating the sensor’s behaviour during exposure to each VOC and subsequent $ \mathrm{N_2} $ purging.
Water exhibits the weakest response, confirming that only a trace amount is adsorbed in the framework.  
The experiments show that VOC absorption is favoured over H$_2$O, in agreement with earlier reports by Pandey \emph{et al.} and Sann \emph{et al.}\cite{pandey2024fabrication,sann2018highly}. In contrast, ZIF-8 shows its strongest response\cite{zhang2013adsorption} toward 2-butanone and the series of C$_1$–C$_4$ alcohols.

Toluene induces only a moderate change in capacitance. Although its kinetic diameter exceeds the $3.4\,\text{\AA}$ aperture of ZIF-8, framework “gate-opening’’ might permit uptake, as demonstrated by Khudozhitkov \emph{et al.}\cite{khudozhitkov2019mobility}  
The reduced signal could originate from toluene’s very low dielectric constant ($2.38$ versus $18\text{–}78$ and its modest electronic polarizability compared to that of the other VOCs\cite{akerlof1932dielectric}). The observed differences in sensor response can arise from a combination of dielectric matching, pore-size exclusion, and guest–host interactions, as detailed in Tables~\ref{tab:VOC_properties} and \ref{tab:materials} and as discussed in Sec.~\ref{sec:voc-water-sensing-mechanism}.



The sensor response ($\Delta \text{ADC}$) of both inks towards all VOCs and water is shown in Fig.~\ref{fig:hydrophobicity-test}(e), demonstrating distinct interaction behaviors between ZIF-8 and UV ink. 
The results indicate that ZIF-8 ink exhibits a significantly stronger response for all VOCs except toluene, compared to UV ink. This behavior suggests that UV ink interacts more weakly with VOC molecules, likely due to the absence of microporosity compared to ZIF-8. The enhanced response of ZIF-8 ink to ketones (such as 2-butanone) and alcohols (including ethanol, methanol, 2-propanol, and 1-butanol) suggests its preferential adsorption for molecules with these functional groups.
The ratios of responses to a gas across the different inks in Fig.~\ref{fig:hydrophobicity-test}(f) provide more detailed insight into the sensor's selectivity. This highlights their relative interactions with different VOCs. The ZIF-8/UV ratio peaks for 2-butanone, whereas the reciprocal ratio is largest for toluene, indicating that the UV ink responds more strongly to it than the ZIF-8 ink.

\CHANGE{Thus, in Fig.~\ref{fig:hydrophobicity-test}(f) selectivity is expressed as the ratios
$\Delta\mathrm{ADC}_{\text{ZIF-8 ink}}/\Delta\mathrm{ADC}_{\text{UV ink}}$ and $\Delta\mathrm{ADC}_{\text{UV ink}}/\Delta\mathrm{ADC}_{\text{ZIF-8 ink}}$ for each VOC. The high ZIF-8/UV ratios for 2-butanone and the C$_1$--C$_4$ alcohols reflect ZIF-8's stronger response to these polar VOCs, consistent with their favourable adsorption and higher dielectric constants. Conversely, the largest UV/ZIF-8 ratio is observed for toluene, confirming that the neat UV resin is more sensitive to toluene than the ZIF-8 ink. This can be attributed to the non-porous, carbon-rich polymer matrix of the UV ink, which favours sorption of non-polar aromatic molecules, whereas ZIF-8 provides microporous adsorption sites that favour smaller, more polar VOCs.}

\CHANGE{It should be emphasized that neither ZIF-8 nor the UV-curable ink is specific to a single vapour. Rather, each ink exhibits a characteristic response pattern across the VOC panel, and selectivity arises from comparing these patterns in an array context. The
relatively low response of ZIF-8 to water and toluene in Fig.~\ref{fig:hydrophobicity-test}(e) can be understood by considering both uptake and dielectric properties. ZIF-8 is strongly
hydrophobic and its water uptake remains negligible up to about 80\% relative humidity, as shown by the water adsorption isotherm in Fig.~\ref{fig:ZIF8-adsorption}(c); at the $\sim$20\%~RH corresponding to 6000~ppm H$_2$O in our experiments, only a trace amount of water is adsorbed, leading to a very small capacitance change. For toluene, the combination of limited uptake in the $\sim$3.4--4~\AA\ windows of ZIF-8, as discussed in Sec.~\ref{sec:voc-water-sensing-mechanism}, and the very low dielectric constant of toluene ($\varepsilon_r \approx 2.38$) results in a modest signal. In contrast, the UV-curable polymer matrix is non-porous but highly carbon-rich, and can swell and absorb toluene
more effectively, which explains its larger response to toluene. Other VOCs in the test set exhibit both higher uptake in ZIF-8 and higher dielectric constants, producing the stronger signals observed in Fig.~\ref{fig:hydrophobicity-test}(e).}

\begin{figure}[htpb]
    \centering
    \includegraphics[width=0.9\textwidth]{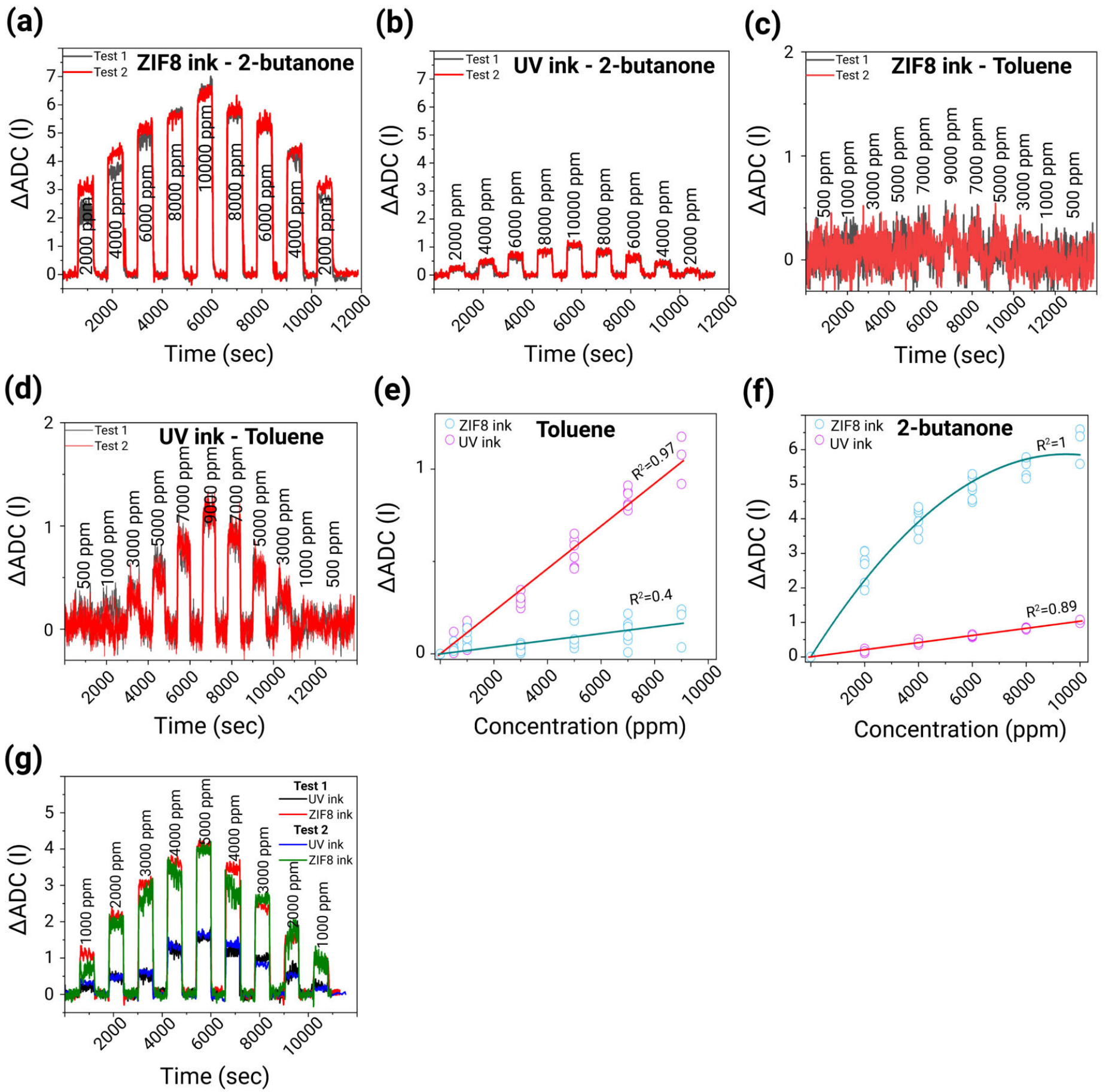}
    \caption{
        \textbf{Sensor calibration for single VOC}   
        (a) ZIF-8 ink-functionalised area response to 2-butanone,  
        (b) UV ink-functionalised area response to 2-butanone,  
        (c) ZIF-8 ink functionalised area response to toluene,  
        (d) UV ink-functionalised area response to toluene.  
        (e–f) Sensor calibration and fits:  
        (e) Linear and quadratic fit of the sensor responses of the  ZIF-8 and UV ink functionalised areas resp. to 2-butanone,  
        (f) linear fits of the responses of the ZIF-8 and UV ink functionalised areas to toluene.  
        (g) Sensor response of UV and ZIF-8 ink-functionalised areas to mixtures of toluene and 2-butanone gas in varying concentrations, where concentrations of both VOCs are equal, as indicated in the graph. 
        }
    \label{fig:sensor-responses}
\end{figure}

\subsubsection{Sensor calibration for single VOC gas}

Based on the initial multi-gas screening, we selected 2-butanone and toluene for subsequent analyses. A typical measurement is shown in Fig.~\ref{fig:sensor-responses}(a), where the concentration of 2-butanone is stepwise increased and decreased stepwise between 2000 and 10000 ppm and, in between, set to 0 ppm. To check the reproducibility of the measurement, each test is performed twice, as indicated by the black and red lines in the graph.

To enable quantification of gas concentrations, we first perform a calibration of the pixelated sensor for the two pure gases, 2-butanone and toluene. The response of the sensors functionalized with ZIF-8 ink and UV ink to both VOCs is characterized as outlined in the previous subsection. The results are shown in Figs.~\ref{fig:sensor-responses}(a–f). The $\Delta \mathrm{ADC}$ values from these measurements are averaged over multiple pixels. Two tests (Test 1 and Test 2) are performed, indicated by black and red lines in Figs.~\ref{fig:sensor-responses}(a–d), resulting in four values per gas concentration for each ink–VOC combination, which are plotted as circles in Figs.~\ref{fig:sensor-responses}(e-f).  

Three of the four datasets exhibit a linear trend and were fitted with a linear model,  
$\Delta \text{ADC}_{\text{ink,VOC}} = c_{\text{ink,VOC},1} \cdot C_{\text{VOC}}$.  
The response of ZIF-8 to 2-butanone shows a saturation effect and was therefore fitted with a quadratic model, 
$\Delta \text{ADC}_{\text{ZIF,butanone}} = c_{\text{ZIF,Bu},1} \cdot C_{\text{butanone}} + c_{\text{ZIF,Bu},2} \cdot C_{\text{butanone}}^2$.  
All fits were performed using the least-squares regression method. The resulting fit parameters from this calibration procedure are summarized in Table~\ref{tab:fitting_coefficients}, which shows in particular that the response of the ZIF-8 ink to butanone is much stronger than its response to toluene, providing selectivity.

\begin{table}[htbp]
    \centering
    \caption{Extracted fitting coefficients for UV and ZIF-8 ink responses. 
    Concentrations are given in ppm. Since $\Delta \text{ADC}(I)$ is dimensionless, 
    the linear coefficients $c_{1}$ have units of 1/ppm and the quadratic coefficient 
    $c_{2}$ has units of 1/ppm$^{2}$.}
    \label{tab:fitting_coefficients}
    \begin{tabular}{|c|c|l|}
        \hline
        \textbf{Coefficient} & \textbf{Extracted Value} & \textbf{Description (Units)} \\ 
        \hline
        $c_{\text{UV,To},1}$ & $1.153 \times 10^{-4}$ & Sensitivity of UV ink to toluene (1/ppm) \\ 
        $c_{\text{UV,Bu},1}$ & $1.031 \times 10^{-4}$ & Sensitivity of UV ink to butanone (1/ppm) \\ 
        $c_{\text{ZIF,To},1}$ & $1.830 \times 10^{-5}$ & Sensitivity of ZIF-8 ink to toluene (1/ppm) \\ 
        $c_{\text{ZIF,Bu},1}$ & $1.238 \times 10^{-3}$ & Linear sensitivity of ZIF-8 ink to butanone (1/ppm) \\ 
        $c_{\text{ZIF,Bu},2}$ & $-6.522 \times 10^{-8}$ & Quadratic sensitivity of ZIF-8 ink to butanone (1/ppm$^2$) \\ 
        \hline
    \end{tabular}
\end{table}

Thus, the E-nose is calibrated for its response to two pure gases, 2-butanone and toluene. The measured $\Delta \text{ADC}$ values for the two inks provide reproducible and distinguishable responses for each VOC, enabling quantitative calibration curves as shown in Figs.~\ref{fig:sensor-responses}(e-f). These calibration results form the experimental basis for distinguishing different analytes.

To further evaluate the response of the pixelated capacitive sensor under binary gas exposure, we exposed it to mixtures of toluene and 2-butanone at equal concentrations in each cycle. The mixtures were applied stepwise from 1000 ppm of each gas up to 5000 ppm of each gas, as shown in Fig.~\ref{fig:sensor-responses}(g). The resulting $\Delta \mathrm{ADC}$ values for both inks are plotted, with UV ink shown in black and blue lines and ZIF-8 ink in red and green lines for two independent test runs. The reproducibility between the repeated exposures confirms the stability of the sensor response to binary VOC mixtures when the two components are present at the same concentration.

\CHANGE{Besides equimolar mixtures, we also performed a preliminary experiment in which the total VOC concentration was held constant while the relative fractions of toluene and 2-butanone were varied (see Fig.~\ref{fig:mix_dynamic}). In this case, the ZIF-8 and UV-ink regions respond in opposite directions as the mixing ratio changes, confirming that the paired responses carry information on the composition of non-equimolar binary mixtures.}

\section{Discussion}
\label{sec:Future}

In this work, we present an E-nose with chip-integrated capacitive readout electronics that can be realized in conventional CMOS technology. Depending on the required gas sensing capabilities, the sensor area is functionalized by inkjet printing with dedicated inks. As a proof-of-principle of using the capacitive pixelated E-nose for sensing, we demonstrate the capability of the chip for detecting the response to pure gases and to binary mixtures of VOCs. 

For capacitive MOF sensors, dry N$_2$ is an appropriate proxy for dry synthetic air with respect to VOC responses; observed differences between dry N$_2$ and ambient air stem primarily from humidity and background VOCs. The MOFs employed here exhibit low water uptake, limiting humidity cross-sensitivity over the ranges tested.

From these experiments, several key performance indicators can be inferred.
First of all, we note from graphs like \ref{fig:sensor-responses}(c) that the CMOS chip is able to distinguish very small capacitance changes of around $\Delta \mathrm{ADC}$=0.1, corresponding to 0.3 aF per sensor element. Noting that the simulated base capacitance of a functionalized sensor element is roughly 0.5 fF, this enables detecting changes in the effective dielectric constant of the order of 0.1\%, contributing to a high sensitivity.

\CHANGE{In our experiments, we demonstrated reproducible responses for toluene and 2-butanone in the 100--1000~ppm range. Using the rms baseline noise of the averaged
$\Delta\mathrm{ADC}$ signal in dry N$_2$ ($\mathrm{rms}_{\mathrm{noise}} \approx 0.03$~ADC) and
the slopes of the calibration curves in Figs.~7(e,f), and applying the standard $3\sigma$ criterion $\mathrm{LOD} = 3\,\mathrm{rms}_{\mathrm{noise}}/S$, we estimated ink-
 and VOC-specific limits of detection (see in~ Table~\ref{tab:LOD}). For the most responsive
combination, ZIF-8 exposed to 2-butanone, the resulting LOD is approximately $7.3\times10^{1}$~ppm. For the UV ink, the estimated LODs are about $7.8\times10^{2}$~ppm for toluene and $8.7\times10^{2}$~ppm for 2-butanone, whereas
the weak response of ZIF-8 to toluene leads to a higher LOD of $\sim 4.9\times10^{3}$~ppm. These values reflect the trade-off between sensitivity and selectivity for different
ink--VOC pairs and are consistent with the relative slopes observed in Figs.~7(e,f).}

We note, however, that we foresee substantial possibilities for improving the detection limit. By engineering the electrode configuration and by improving ink composition, deposition strategy, and porosity, while also reducing sensor noise levels, it might well be possible to bring detection limits down to the 1-10 ppm range.
A special point of notice is the reproducibility and fast response of the current sensor. As is seen in graphs like \ref{fig:sensor-responses}(c), the sensor returns rapidly (within $\sim$100 s) and reproducibly to the initial $\Delta \mathrm{ADC}$=0 value after removing the VOC gas. This response time depends both on the sensor itself and on the time it takes to stabilize the gas concentration at the sensor surface after removing the VOC. The actual sensor response might even be faster than 100 seconds, since at the operating flow rates, the time for replacing the gas in the tubes after changing the concentration setting is of the order of a minute and thus might be limiting the response in our experiments. Since the readout electronics are very fast, we expect response times to ultimately be limited by diffusion and effusion rates of gases in the ink functionalised layers. Response times might thus be further reduced by using thinner functionalisation layers, although this will probably reduce sensitivity. 

The reproducibility and absence of large drifts of the sensor indicate that the gases cause no permanent chemical or physical changes in the functionalised layers. Most likely, the observed signals are therefore a result of the diffusion and effusion of the gas into the materials of which the functionalised layers consist, onto the external surfaces of the materials, but also into porous ZIF-8 and absorbed into the deposited polymer of the UV-ink. The gas that enters the bulk and/or surface of the functionalisation layer, will cause a change in the effective dielectric constant. It is known \cite{Khalaj2023, Tanaka2015} that both toluene and 2-butanone absorb well to ZIF-8 particles. While 2-butanone has a high relative dielectric constant of approximately 18.5, toluene has a much lower dielectric constant of 2.38. The high absorption of 2-butanone in combination with its high dielectric constant can explain the high response in Fig.~\ref{fig:sensor-responses}(h) to this gas, especially of the porous ZIF-8 in the ink. The response of the sensor to toluene gas is much weaker, and surprisingly, the response of the ZIF-8 functionalised layer to toluene is weaker than the response of the layer based on UV ink in Fig.~\ref{fig:sensor-responses}(g). 

While a hypothetical explanation could be displacement of residual H$_2$O in ZIF-8 by toluene, this is unlikely because ZIF-8 is highly hydrophobic and the measurements were performed under dry gas conditions. Instead, the significantly higher response of the UV-ink functionalised layer suggests that toluene is absorbed into the acrylate-based polymer matrix, potentially even swelling the polymer, which would allow higher uptake compared to ZIF-8, which cannot swell. Furthermore, the UV-ink shows a stronger response to toluene than to 2-butanone, despite the latter having a higher dielectric constant. This indicates that the absorption affinity is governed by chemical interactions: the high aliphatic carbon content of the UV-ink polymer matrix favors uptake of the apolar toluene. Elemental analysis by total EDX intensity-image (97.5\% carbon, see SI) supports this interpretation, consistent with preferential sorption of toluene over 2-butanone in the UV-ink layer.

This opposite response of the inks towards toluene and 2-butanone, with UV ink having the highest response to toluene and ZIF-8 ink having the highest response to 2-butanone, is beneficial for using these inks for determining the gas concentration in a mixture of the two gases. In general, for good selectivity of E-noses, it is important to find combinations of functionalization materials that are specifically sensitive to some of the gases compared to the other gases. 


\label{eq:additivity2}
\CHANGE{The pixelated architecture with 1024 electrodes per chip is designed not only to improve signal-to-noise ratio by spatial averaging, but also to enable a high degree of chemical diversity on a single CMOS platform. In the present proof-of-principle, we use four functional layers (three MOF-based inks and one UV-curable polymer), but the same inkjet-printing approach can, in principle, be extended to many more materials. Selectivity
can be modulated by combining MOFs with different pore sizes, topologies, and surface chemistries, by varying the MOF loading and composition of the polymer matrix, and by printing different inks in spatially separated microdomains on each sensor matrix. With a conservative estimate of a $\sim$50~\si{\micro\meter} droplet footprint, a single $\sim$400~$\times$~500~\si{\micro\meter\squared}
matrix can host on the order of 80 distinct functionalization regions, so that using all three matrices allows for up to $\sim$240 different inks on one chip.}

\CHANGE{Using many partially-specific sensors in parallel increases the dimensionality of the odour fingerprint and thereby improves discrimination of complex gas mixtures, analogous to the
way the mammalian olfactory system benefits from a large repertoire of broadly responsive olfactory receptors. At the same time, the time-division multiplexed readout architecture
ensures that power and data-rate requirements remain manageable: all 1024 electrodes in a matrix share the same front-end and ADC and are read out sequentially, so the power
consumption does not scale linearly with the number of pixels. In our implementation, the
full array can be scanned every 0.3–0.5~s, which is compatible with the characteristic
adsorption/desorption times of the VOCs studied.}

\CHANGE{The present study focuses on the CMOS capacitive E-nose platform itself, its functionalization, and calibration under controlled laboratory conditions. 
A further extension will be to upgrade the gas-mixing manifold with additional independently controlled VOC lines so that ternary and more complex mixtures (including
interferents such as ethanol) can be investigated systematically and used to train multivariate models for anti-interference and decoupling of multi-component gas mixtures. In addition, translating this platform to specific application scenarios, such as greenhouse VOC monitoring or food spoilage detection, will require dedicated sampling and packaging solutions that account for dust, humidity, temperature variations, and gas-flow conditioning, including appropriate pre-filters. Designing and validating such application-specific modules, and conducting long-term field trials, fall outside the scope of this work and will be the subject of future studies, in which the chip will be integrated into complete sensor nodes and tested under realistic environmental conditions.}

\section{Conclusion}

We present a pixelated CMOS capacitive sensor chip that is functionalized by inkjet printing with MOF and polymer inks. The E-nose is shown to detect volatile organic compounds, with a response that can be controlled by the ink composition. A dedicated measurement methodology and analysis software are developed to process the simultaneous and continuous response of all 1024 pixel sensors and improve capacitance resolution by averaging over multiple pixels in selected functionalized regions. 
\CHANGE{The sensor chip shows a stable and fast response to changes in VOC concentrations
down to 100~ppm within $\sim$100~s under our experimental conditions.} It was found that the ZIF-8 and pure UV-curable ink were the most hydrophobic compared to MIL-101(Cr) and MIL-140A. The complementary selectivity of the inks enables discrimination between 2-butanone and toluene exposure. After calibration in pure VOCs of known concentration, reproducible responses were also observed for controlled binary mixtures of the two gases. 
The small size, low cost, and high performance of the CMOS readout, together with the flexibility of inkjet-based functionalization, highlight the potential of this approach for further development into a multi-gas sensing platform. A remaining challenge is the synthesis and selection of inks with both high sensitivity and selectivity to targeted gases. With further optimization, the presented platform could be applied in safety monitoring, agriculture, the food industry, and robotics.

\begin{acknowledgement}
M.A.B.-M.K., P.G.S., and M.K.G. acknowledge support from the Dutch government as part of the National Growth Fund programme NXTGEN Hightech, project Agrifood 11: Greenhouse Horticulture - Digital Twin. M.A.v.d.V, C.H., and F.W. acknowledge support from the research programme Nationale Wetenschapsagenda – Onderzoek op Routes door Consortia (NWA-ORC) 2020/21, which is (partly) financed by the Dutch Research Council (NWO) under number NWA.1389.20.123. P.G.S. acknowledges funding from the European Union under the Horizon Europe Programme, Grant Agreement No. 101136388.
\end{acknowledgement}

\newpage


\begin{suppinfo}
\end{suppinfo}

\newpage

\newpage

\renewcommand{\theequation}{S\arabic{equation}}
\renewcommand{\thesection}{S\arabic{section}}
\renewcommand{\thepage}{S\arabic{page}}
\renewcommand{\thetable}{S\arabic{table}}
\renewcommand{\thefigure}{S\arabic{figure}}

\setcounter{section}{0}
\setcounter{figure}{0}
\setcounter{equation}{0}
\setcounter{table}{0}
\setcounter{page}{1}

\section*{Supporting Information} 

\section{Antoine Equation and Vapor Pressure Calculations}
\label{sec:antoine-equation}

To determine the saturated vapor pressure $P_s$ of the VOC gases under study in the bubbler used for the gas sensing setup, we use the Antoine equation:
\begin{equation}
    \log_{10} \left( \frac{P_s}{P_0} \right) = A - \frac{B}{C + T}
    \label{eq:antoine}
\end{equation}

Where:
\begin{itemize}
    \item \( P_s \) is the saturated vapor pressure of the VOC (\text{mmHg}).
    \item \( P_0 \) is the reference pressure, typically taken as the ambient pressure \((P_0 = 760 \text{ mmHg})\).
    \item \( A \) is an Antoine constant, typically unitless.
    \item \( B \) is an Antoine constant with units of temperature \((\degree C)\).
    \item \( C \) is an Antoine constant with units of temperature \((\degree C)\).
    \item \( T \) is the temperature in degrees Celsius \(\degree C\).
\end{itemize}
 
In Equation~\ref{eq:antoine}, the Antoine equation expresses the relationship between vapor pressure and temperature for a given compound.
The Antoine equation was used to calculate the saturated vapor pressure \( P_s \) of the selected VOCs. The constants \(A\), \(B\), and \(C\) used in Equation~\ref{eq:antoine} for these VOCs are listed in Table~\ref{tab:antoine-constants}. These values were obtained from references \cite{de2012physical, hoa2009nanowire, nguyen2011meso} and are valid for the temperature range used in the experiments.

\begin{table}[htbp]
\centering
\caption{Antoine constants used for selected VOCs.}
\label{tab:antoine-constants}
\begin{tabular}{lccc}
\toprule
\textbf{VOC} & \textbf{A} & \textbf{B} & \textbf{C} \\
\midrule
Ethanol       & 8.13484 & 1662.48 & 238.131 \\
2-Propanol    & 8.87829 & 2010.33 & 252.636 \\
Methanol      & 8.09126 & 1582.91 & 239.096 \\
Water         & 8.07131 & 1730.63 & 233.426 \\
1-Butanol     & 7.62121 & 1543.89 & 208.029 \\
toluene       & 7.13620 & 1457.29 & 231.827 \\
2-butanone (MEK) & 7.29427 & 1400.37 & 237.655 \\
\bottomrule
\end{tabular}
\end{table}

\CHANGE{
\begin{table}[htbp]
\centering
\caption{Structural and chemical characteristics of the investigated MOFs.}
\label{tab:mof_structure}
\begin{tabular}{lccc}
\toprule
MOF          & Chemical formula                       & Window size (\AA) & Cage size (\AA) \\
\midrule
ZIF-8        & Zn(mIm)$_2$                             & 3.4               & 11.6            \\
MIL-140A     & ZrO(BDC)                                & 3.2               & n.a.            \\
MIL-101(Cr)  & Cr$_3$(O)Cl(BDC)$_3$(H$_2$O)$_2$        & 12, 16 $\times$ 14.5 & 29, 34      \\
\bottomrule
\end{tabular}
\end{table}
}

\section{Characterization of MOF Nanoparticles}
\label{sec:mof-nanoparticles-characterization}
This section presents supplementary information on the characterization of MOFs and the CMOS chip.
\subsection{Structural and Adsorption Analysis of ZIF-8 Nanoparticles}
\subsection{Structural Characterization}

This section provides detailed structural and adsorption analyses of the synthesized ZIF-8 nanoparticles. The following sections outline various characterization methods and results that complement the main article.

\begin{figure}[htbp]
    \centering
    \includegraphics[width=0.6\textwidth]{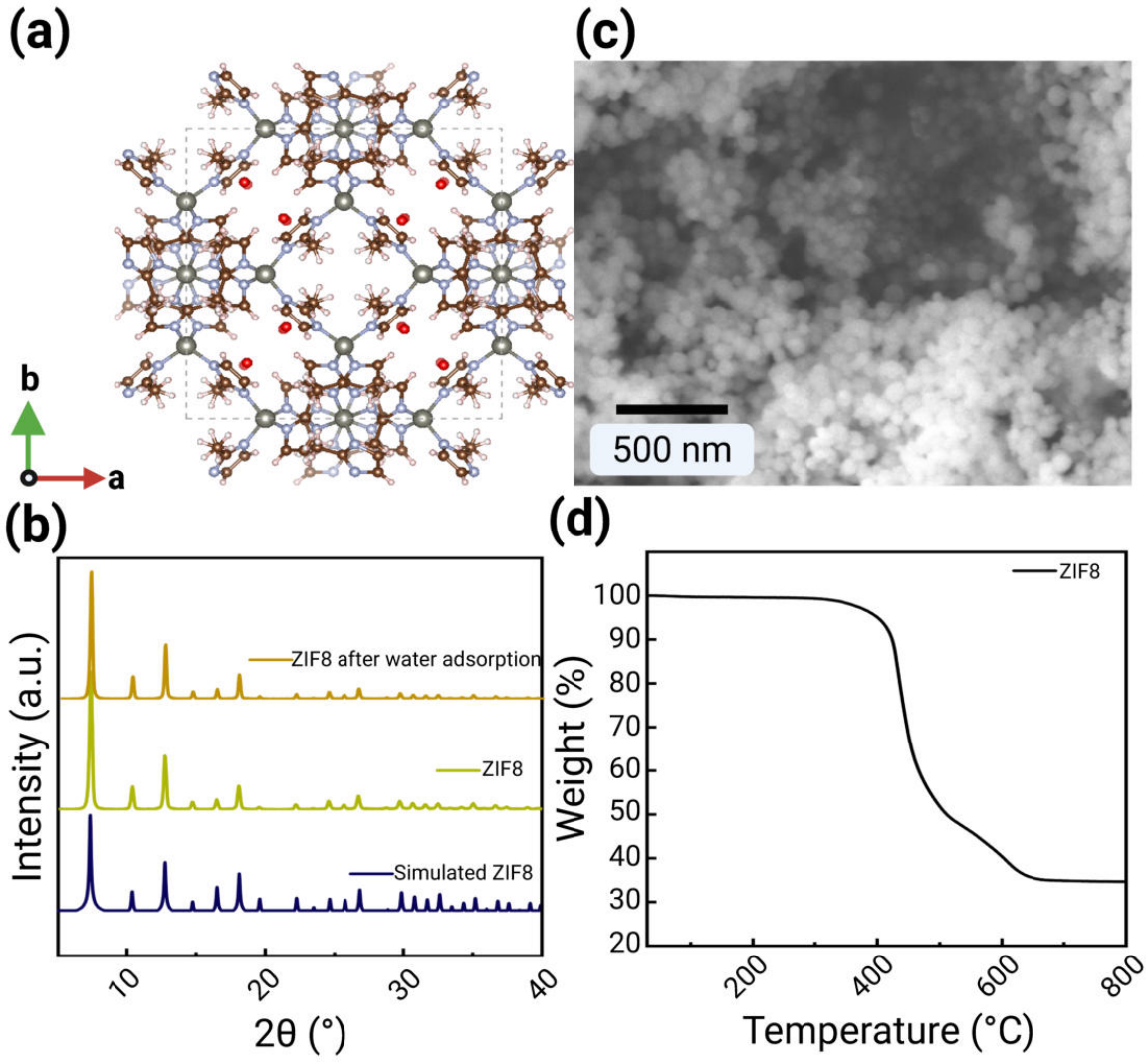}
    \caption{
        \textbf{Structural characterization of ZIF-8.}  
        (a) Simulated porous structure of ZIF-8 viewing along the c-axis, grey dashed lines represent the unit cell.
        (b) PXRD of synthesized ZIF-8, ZIF-8 after water adsorption, and simulated patterns. 
        (c) SEM image of ZIF-8 nanoparticles.
        (d) TGA analysis demonstrating thermal stability up to 400$^\circ$C.
    }
    \label{fig:ZIF8-structure}
\end{figure}

The structural properties of the synthesized ZIF-8 nanoparticles were analyzed using FESEM, Powder X-ray diffraction (PXRD), and thermogravimetric analysis (TGA), as presented in Fig.~\ref{fig:ZIF8-structure}.

Fig.~\ref{fig:ZIF8-structure}(a) shows the crystal structure of ZIF-8, illustrating its well-defined microporous network. This crystal structure highlights the topological arrangement of Zn metal nodes and imidazolate linkers, which contribute to the high surface area and adsorption capacity of ZIF-8.
The structure and crystallinity of the synthesized ZIF-8 sample were verified using PXRD, as shown in Fig.~\ref{fig:ZIF8-structure}(b). The diffraction peaks match well with simulated patterns, confirming the highly crystalline nature of ZIF-8 and the absence of secondary phases. Moreover, the PXRD pattern of ZIF-8 after water adsorption remained the same, indicating the water(vapor) stability of the structure.
The morphology of the synthesized ZIF-8 nanoparticles is confirmed by the SEM image in Fig.~\ref{fig:ZIF8-structure}(c). The nanoparticles exhibit a uniform octahedral shape with an average size of approximately 100 nm, indicating well-controlled synthesis conditions. This uniformity ensures a smooth inkjet printing process without nozzle clogging.
Finally, thermogravimetric analysis (TGA) was conducted to evaluate the thermal stability of ZIF-8, and the results are shown in Fig.~\ref{fig:ZIF8-structure}d. The weight-loss profile indicates that ZIF-8 maintains structural integrity up to 400$^\circ$C, beyond which decomposition occurs due to ligand breakdown.

\subsection{Adsorption Properties of ZIF-8 Nanoparticles}

\begin{figure}[htbp]
    \centering
    \includegraphics[width=0.7\textwidth]{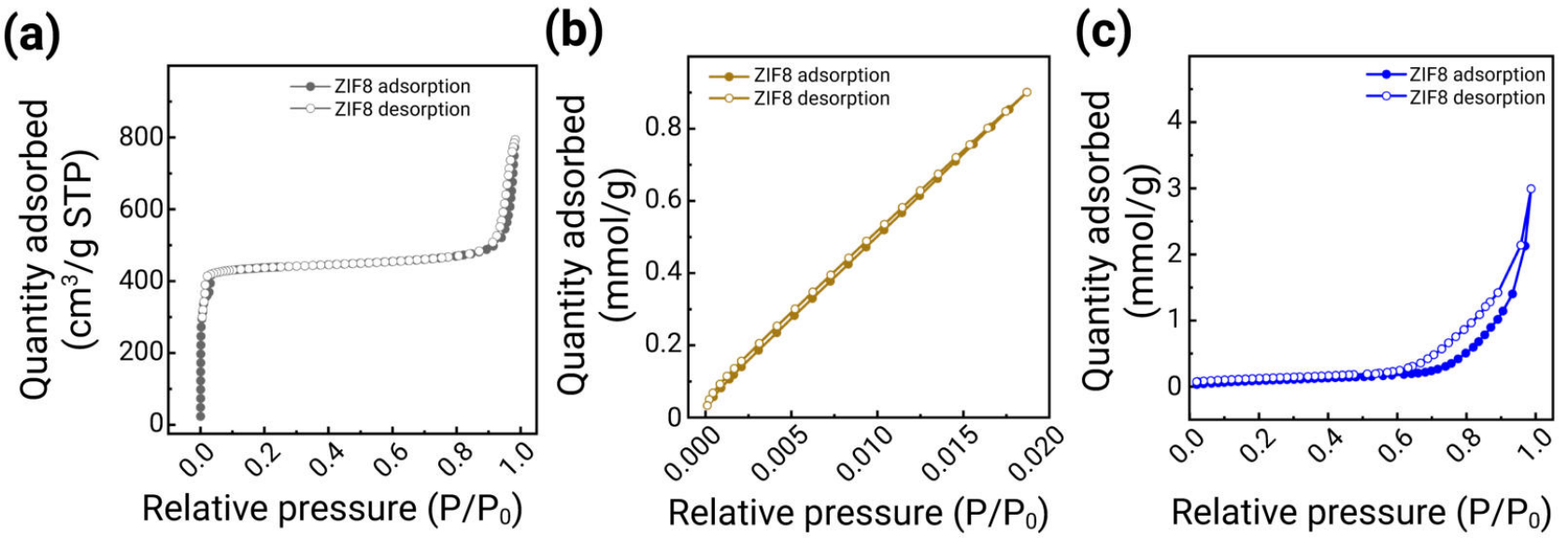}
    \caption{
        \textbf{Adsorption characterization of ZIF-8.}  
        (a) N$_2$ adsorption measured at 77 K.  
        (b) CO$_2$ adsorption measured at 298 K.  
        (c) Water adsorption measured at 293 K.
    }
    \label{fig:ZIF8-adsorption}
\end{figure}

\begin{figure}[ht]
    \centering
    \includegraphics[width=0.4\textwidth]{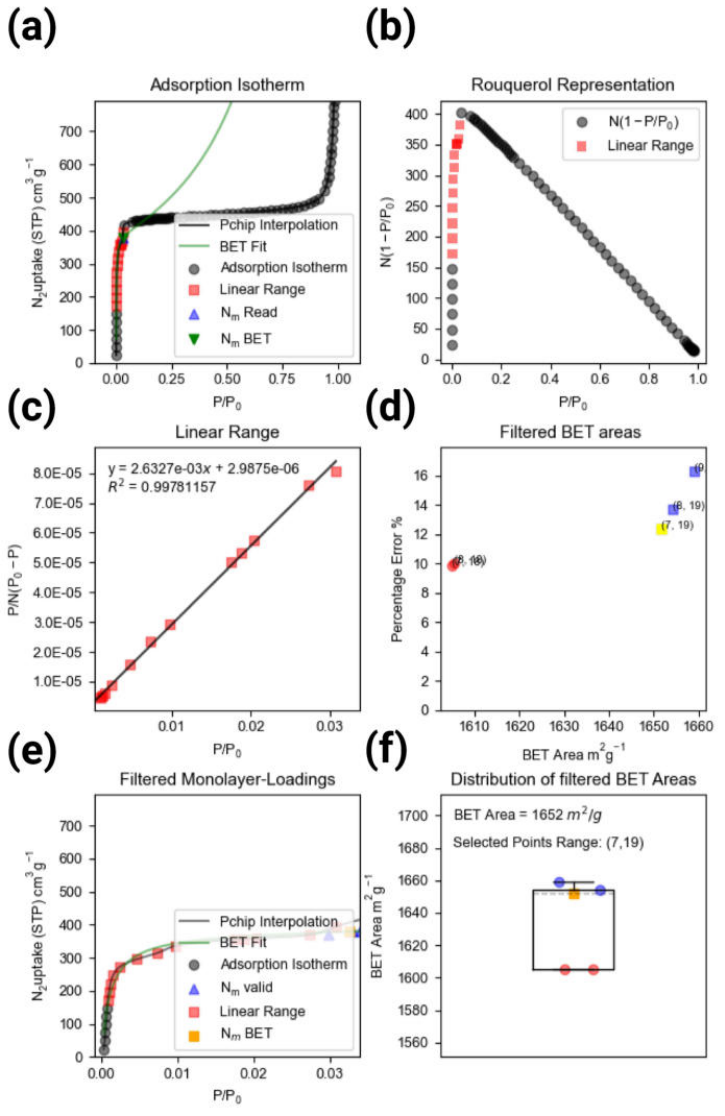}
    \caption{BETSI analysis report for  ZIF-8 (a) adsorption isotherm, (b) Rouquerol representation, (c) linear range, (d) filtered BET areas, and (e) filtered monolayer-loadings, and (f) distribution of filtered BET areas.
    }
    \label{fig:ZIF8-BET}
\end{figure}

Nitrogen adsorption-desorption isotherms were measured at 77 K to evaluate the pore structure and surface area of ZIF-8. 
The isotherms exhibited a typical Type I profile characterized by a steep increase at low relative pressure, indicative of microporous materials, as shown in Fig.~\ref{fig:ZIF8-adsorption}(a). 
The Brunauer-Emmett-Teller (BET) surface area was found to be 1652 m$^2$/g (Fig.~\ref{fig:ZIF8-BET}), which is in line with the reported values~\cite{Allegretto2024}. 


Fig.~\ref{fig:ZIF8-adsorption}(b) illustrates CO$_2$ adsorption at 298 K, the CO$_2$ capacity reaches 0.9 mmol/g at 1.02 bar, and indicates no meaningful adsorption of CO$_2$ at atmospheric concentrations (400 ppm). \CHANGE{Fig.~\ref{fig:ZIF8-adsorption}(c) shows water adsorption–desorption isotherms at 293~K. The water uptake of ZIF-8 remains very small up to \(\sim 80\%\) relative humidity, which we attribute to its methyl-functionalized imidazolate linkers, the saturated Zn–N coordination that leaves no open metal sites, and the hydrophobic internal surface of the cages in combination with relatively narrow 3.4~Å pore windows (Table~S2). These structural features create an unfavourable environment for hydrogen-bonded water clusters and explain the low humidity cross-sensitivity observed in the sensor measurements (Fig.~6(a,b)).}

\CHANGE{Fig.~\ref{fig:ZIF8-adsorption}(c) shows water adsorption–desorption isotherms at 293~K. The water uptake remains very small up to \(\sim 80\%\) relative humidity, confirming the strongly hydrophobic character of ZIF-8 and explaining the negligible humidity cross-sensitivity observed at 6{,}000~ppm H\(_2\)O (\(\sim 20\%\) RH) in the device measurements (Fig.~6(a,b)).}

The results from structural and adsorption characterizations underscore the high-quality synthesis of ZIF-8 nanoparticles, making them suitable for gas sensing and separation applications. Uniform morphology, high thermal and water(vapor) stability, significant surface area, and hydrophobicity further validate their potential for real-world sensing applications.

\subsection{Structural and Adsorption Analysis of MIL-140A Nanoparticles}

This section provides detailed structural and adsorption analyses of the synthesized MIL140A nanosheets.

\begin{figure}[htb]
    \centering
    \includegraphics[width=0.4\textwidth]{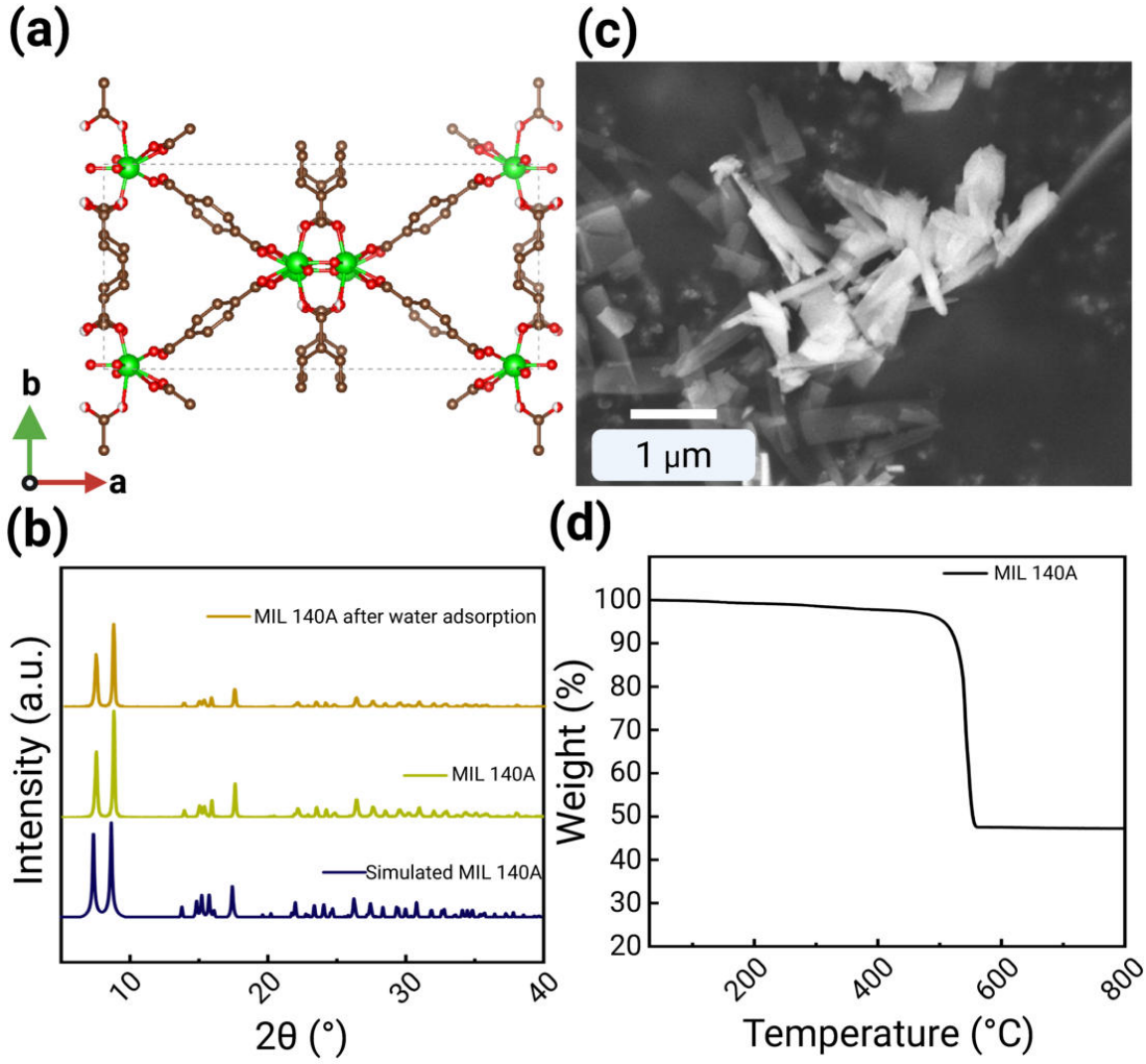}
    \caption{
        \textbf{Structural characterization of MIL-140A.}  
        (a) Simulated porous structure of MIL-140A viewing along the c-axis, grey dashed lines represent the unit cell.
        (b) PXRD patterns comparing the synthesized MIL-140A with its simulated counterpart and post-adsorption sample. 
        (c) SEM image of MIL-140A nanosheets, revealing morphology and uniformity.  
        (d) TGA analysis demonstrating the thermal stability of MIL-140A up to 500$^\circ$C.
    }
    \label{fig:MIL140-structure}
\end{figure}

The structural properties of MIL-140A were analyzed using porosity simulation, SEM, PXRD, and TGA, as presented in Fig.~\ref{fig:MIL140-structure}. Fig.~\ref{fig:MIL140-structure}(a) displays the crystal structure of MIL-140A with a 1D triangular channel. The PXRD patterns shown in Fig.~\ref{fig:MIL140-structure}(b) revealed the successful synthesis of the material compared with the simulated pattern. The XRD pattern of post-water adsorption remained unchanged, suggesting the water stability of MIL-140A. The morphology of MIL-140A is examined via SEM in Fig.~\ref{fig:MIL140-structure}(c). MIL-140A exhibited a nanosheet-like structure with a layer thickness of 10 nm. The layered structure suggests that it has a high external surface area with access to pores, which is beneficial critical for adsorption applications. Finally, TGA in Fig.~\ref{fig:MIL140-structure}(d) demonstrates thermal stability up to 500$^\circ$C, ensuring that MIL-140A retains its structural integrity under high-temperature conditions. These results confirm that MIL-140A exhibits high crystallinity, robust thermal and water stability, and well-defined porosity, making it a promising material for adsorption and gas-sensing applications.

\subsection{Adsorption Properties of MIL-140A Nanoparticles}

\begin{figure}[htb]
    \centering
    \includegraphics[width=0.7\textwidth]{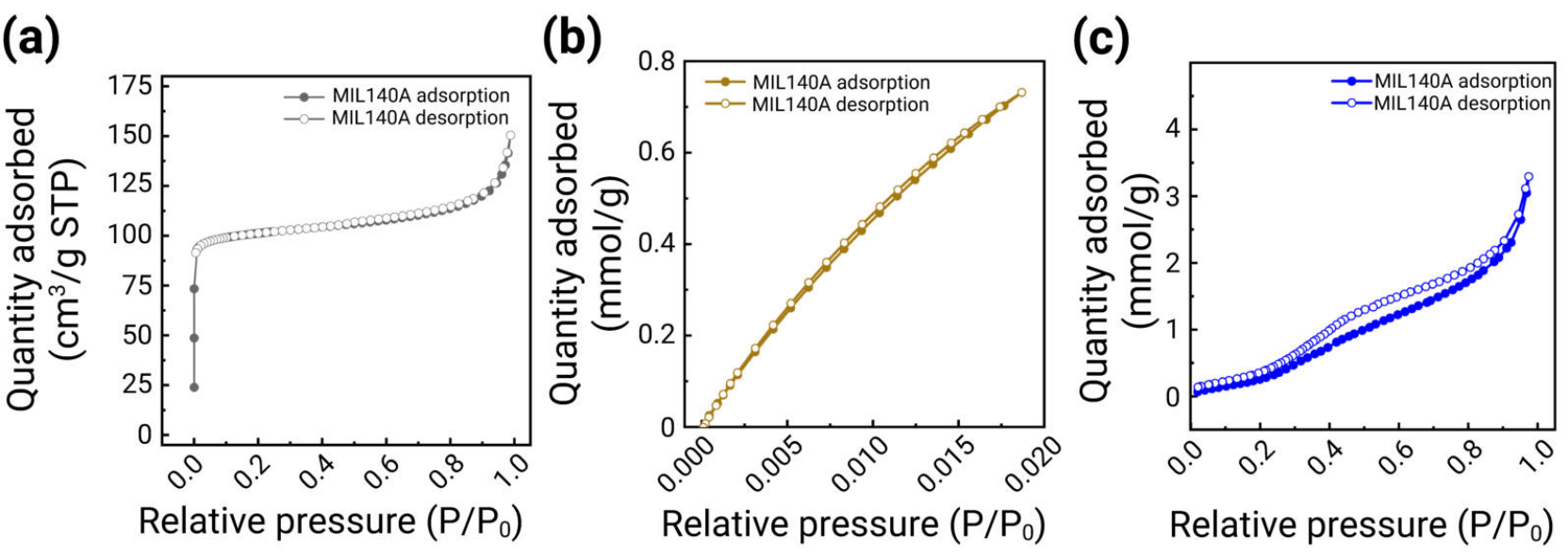}
    \caption{
        \textbf{Adsorption characterization of MIL-140A.}  
        (a) N$_2$ adsorption-desorption isotherm at 77 K, indicating microporous characteristics.  
        (b) CO$_2$ adsorption-desorption isotherm at 298 K, demonstrating CO$_2$ uptake capacity.  
        (c) Water adsorption-desorption isotherm at 293 K.  
    }
    \label{fig:MIL140-adsorption}
\end{figure}

The adsorption properties of MIL-140A were evaluated through N$_2$, CO$_2$, and water adsorption-desorption measurements, as shown in Fig.~\ref{fig:MIL140-adsorption}. Fig.~\ref{fig:MIL140-adsorption}(a) presents the N$_2$ adsorption isotherm at 77 K, which follows a Type I behavior characteristic of microporous materials. The BET surface area of the material is 409 m$^2$/g
, which is consistent with reported values\cite{Schulz2019}.
The steep increase at low relative pressure confirms the presence of small, well-defined pores.  Fig.~\ref{fig:MIL140-adsorption}(b) illustrates CO$_2$ adsorption at 298 K, the CO$_2$ capacity is upto 0.73 mmol/g at 1.02 bar, and indicates no meaningful adsorption of CO$_2$ at atmospheric concentrations (400 ppm). Fig.~\ref{fig:MIL140-adsorption}(c) shows water adsorption at 293 K, revealing the water affinity of MIL-140A. The material showed a gradual water uptake from 20\% relative humidity onward. The uptake behavior suggests moderate hydrophilicity, which is relevant for gas separation applications under humid conditions. The adsorption data confirm that MIL-140A possesses good microporosity and gas uptake capabilities, making it a good candidate for gas storage and separation applications.

\begin{figure}[htbp]
    \centering
    \includegraphics[width=0.4\textwidth]{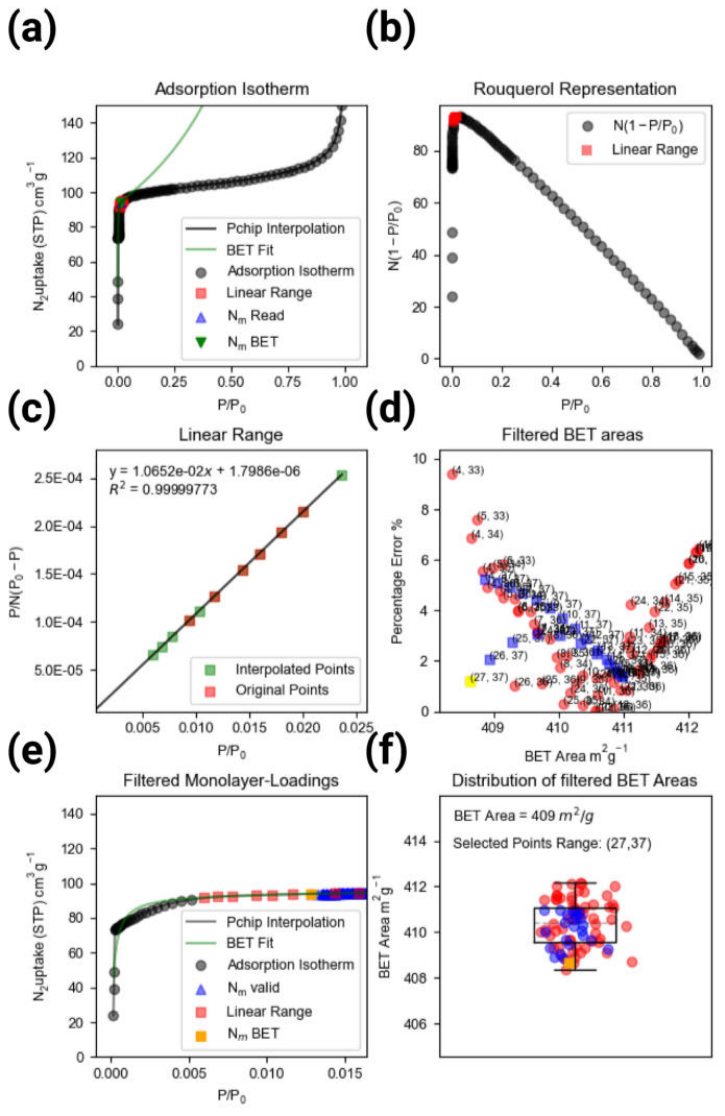}
    \caption{BETSI analysis report for the adsorption isotherm of MIL-140A. (a) adsorption isotherm, (b) Rouquerol representation, (c) linear range, (d) filtered BET areas, (e) filtered monolayer-loadings, and (f) distribution of filtered BET areas.
    }
    \label{fig:MIL-140A-BET}
\end{figure}


\subsection{Structural and Adsorption Characterization of MIL-101(Cr)}

\begin{figure}[htbp]
    \centering
    \includegraphics[width=0.7\textwidth]{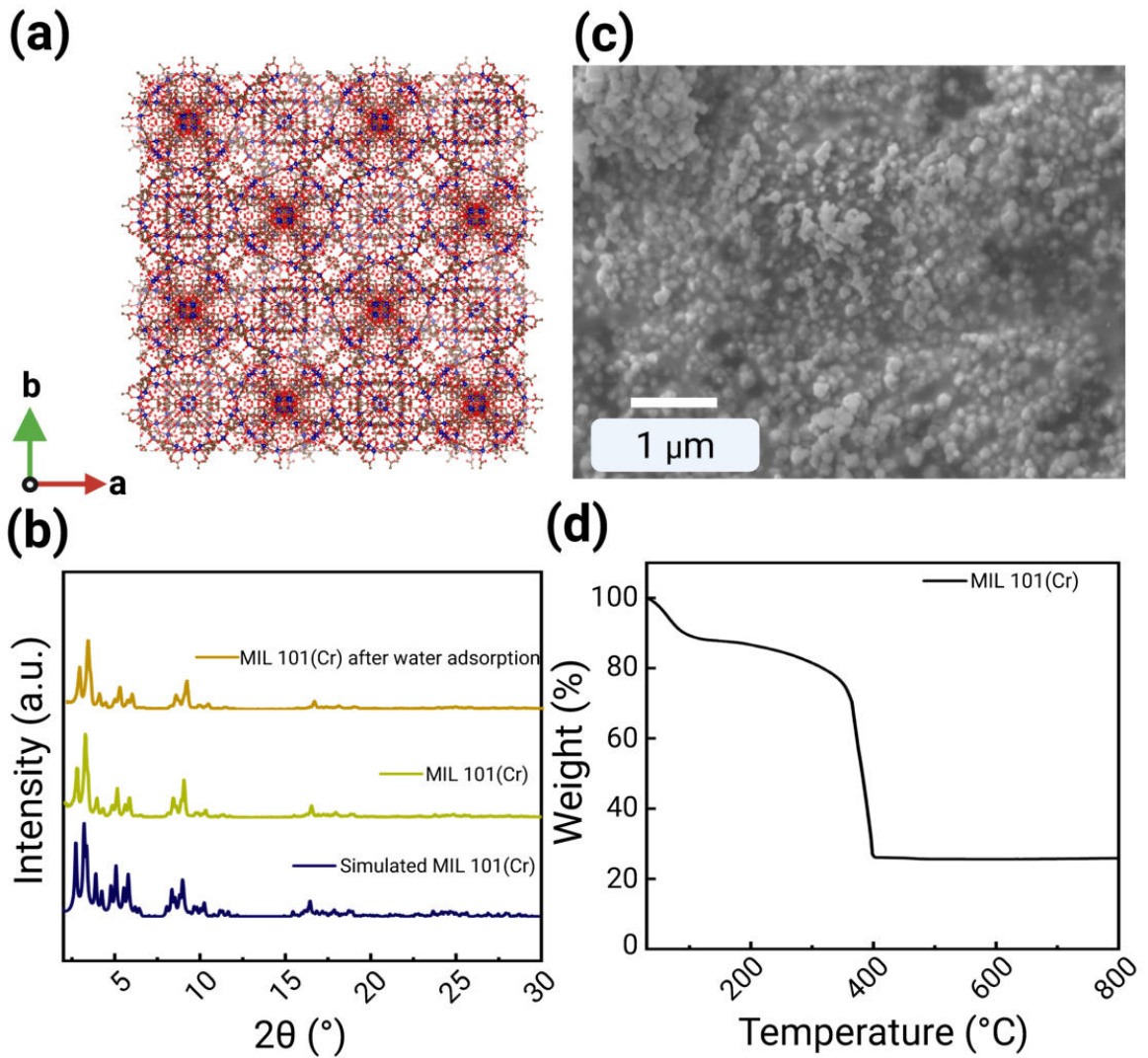}
    \caption{
        \textbf{Structural characterization of MIL-101(Cr).}  
        (a) Simulated porous structure of MIL-101(Cr) viewing along the c-axis, grey dashed lines represent the unit cell. 
        (b) XRD patterns comparing simulated, as-synthesized, and water-adsorbed MIL-101(Cr).
        (c) SEM image showing the morphology of MIL-101(Cr) nanoparticles.     
        (d) TGA showing the thermal stability of MIL-101(Cr).
    }
    \label{fig:MIL101-structure}
\end{figure}

The structural characterization of MIL-101(Cr) was performed using SEM, PXRD, and TGA, as shown in Fig.~\ref{fig:MIL101-structure}. Fig.~\ref{fig:MIL101-structure}(a) shows the crystal structure along the c-axis. Fig.~\ref{fig:MIL101-structure}(b) displays the XRD patterns of simulated, as-synthesized, and post-water adsorption MIL-101(Cr). The characteristic diffraction peaks of the synthesized structure match well with the simulated pattern, confirming the high crystallinity of the material. Fig.~\ref{fig:MIL101-structure}(c) presents an SEM image of MIL-101(Cr) nanoparticles around 100 nm, revealing a well-defined morphology with an aggregated cluster-like structure. The uniformity and size distribution confirms the good control of the synthesis, especially the particle size, making it suitable for subsequent inkjet printing.

Fig.~\ref{fig:MIL101-structure}(d) shows the TGA profile, indicating that MIL-101(Cr) remains thermally stable up to approximately 350°C, beyond which degradation occurs due to organic linker decomposition.

\begin{figure}[htbp]
    \centering
    \includegraphics[width=0.9\textwidth]{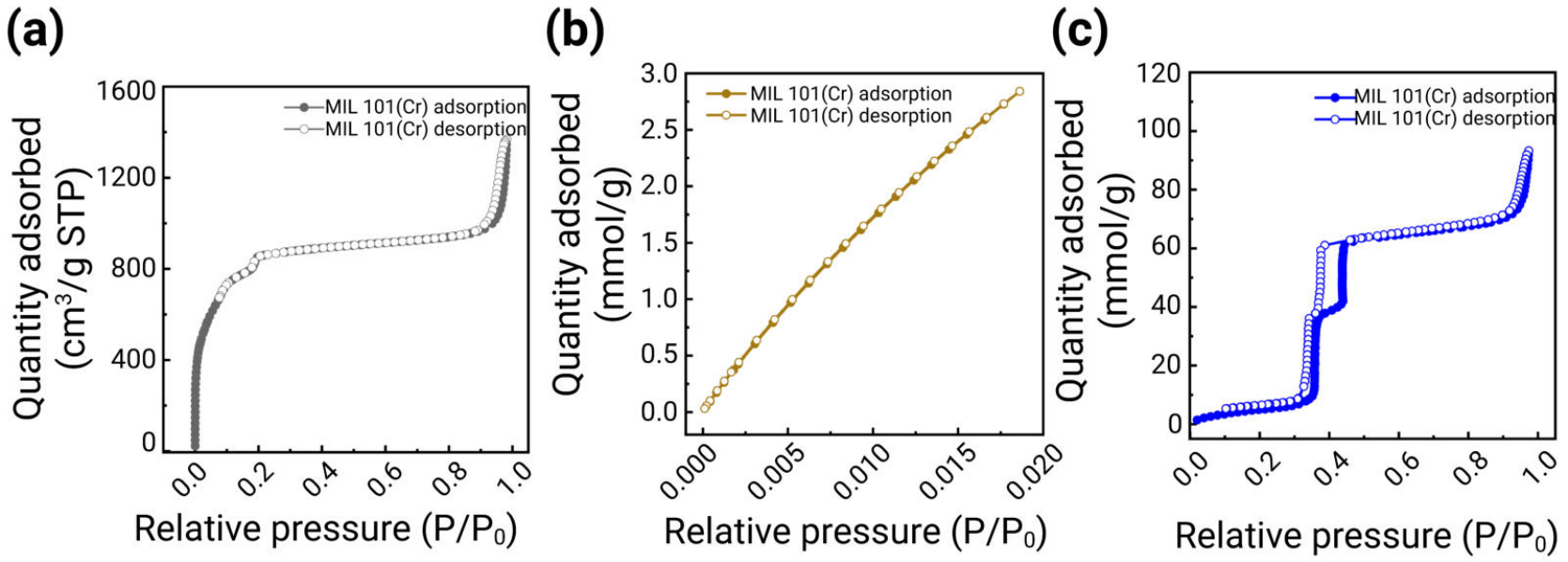}
    \caption{
        \textbf{Adsorption characterization of MIL-101(Cr).}  
        (a) N$_2$ adsorption-desorption isotherms measured at 77 K.  
        (b) CO$_2$ adsorption-desorption isotherms measured at 298 K.  
        (c) Water adsorption-desorption isotherms measured at 293 K.
    }
    \label{fig:MIL101-adsorption}
\end{figure}

The adsorption properties of MIL-101(Cr) were investigated using N$_2$, CO$_2$, and water adsorption-desorption isotherms, as presented in Fig.~\ref{fig:MIL101-adsorption}.
Fig.~\ref{fig:MIL101-adsorption}(a) displays the N$_2$ adsorption isotherm at 77 K. It can be seen that MIL-101(Cr) showed a two-step N$_2$ adsorption, which can be attributed to the 29 \text{\AA} and 34 \text{\AA} cages. The measured BET surface area is found to be 3001 m$^2$/g
, which falls into the decent BET range compared with the reported values \cite{Zorainy2021}, demonstrating that the synthesized MIL-101(Cr) has a high adsorption capacity.
Fig.~\ref{fig:MIL101-adsorption}(b) presents CO$_2$ adsorption isotherms at 298 K, and the CO$_2$ capacity of the material is up to 2.84 mmol/g at 1.02 bar and indicates no meaningful adsorption of CO$_2$ at atmospheric concentrations (400 ppm).
Fig.~\ref{fig:MIL101-adsorption}(c) shows the water adsorption isotherm at 293 K with a two-step water uptake at 0.35 $P/P_0$ and 0.45 $P/P_0$, and hysteresis was observed due to capillary condensation inside the pores\cite{Fei2022}.

\begin{figure}[htbp]
    \centering
    \includegraphics[width=0.4\textwidth]{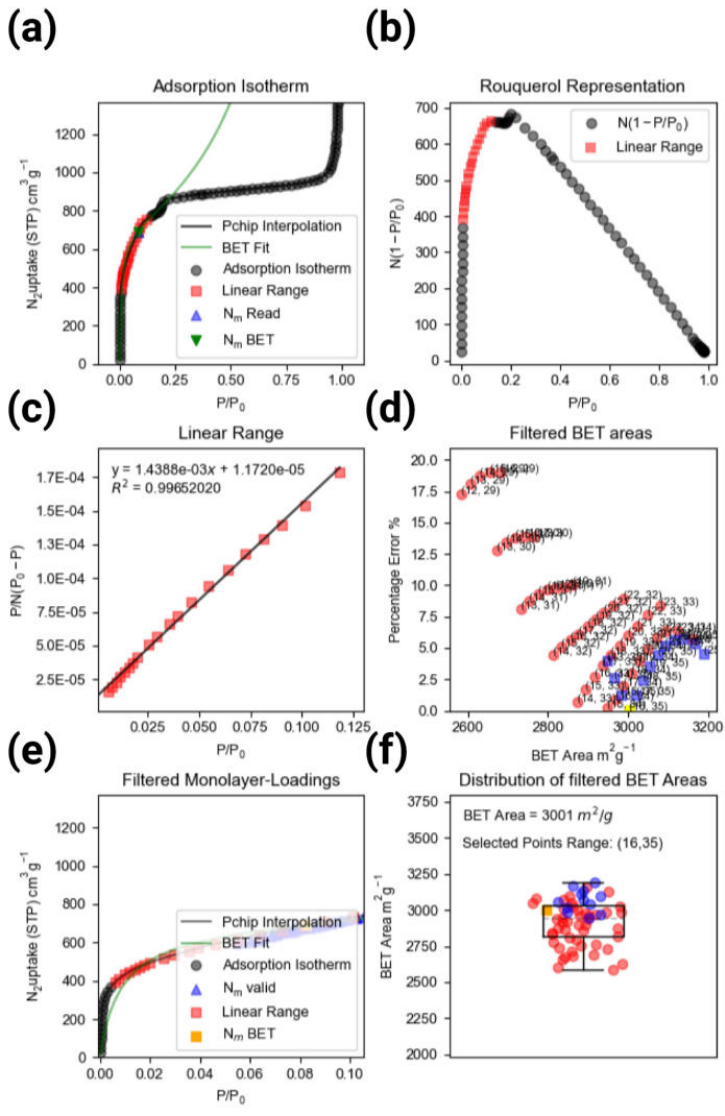}
    \caption{BETSI analysis report for the adsorption isotherm of MIL-101(Cr). (a) adsorption isotherm, (b) Rouquerol representation, (c) linear range, (d) filtered BET areas, (e) filtered monolayer-loadings, and (f) distribution of filtered BET areas.
    }
    \label{fig:MIL-101-BET}
\end{figure}

\subsection{BET calculation on MOF powders}
BET surface area of MOFs was calculated by the BETSI program \cite{osterrieth2022reproducible}. This means that the BET calculation meets the following Rouquerol Criteria: 1) N(1-p/$P_0$) vs P is monotonically increasing; 2) The C constant is positive; 3) The pressure corresponding to the monolayer loading when reported on the isotherm, $N_m$(Read) lies in the fitting range; and 4)$N_m$(Read) and $N_m$(BET), the monolayer loading calculated from BET does not differ by more than 20\%. The BETSI Optimal BET area is chosen as the one with the smallest percentage error under the 4th Rouquerol Criterion, and corresponds to the yellow dot in the filtered BET areas.

\section {Surface and Structural Characterization of Inkjet-Printed ZIF-8 and UV Ink}
\label{sec:surface-structural-characterization}

The surface morphology and structural properties of the inkjet-printed ZIF-8 and UV ink were analyzed using Field Emission Scanning Electron Microscopy (FESEM) and Optical Microscopy. These characterizations confirm the successful deposition of inks and their composition, ensuring reliable sensor functionality.

\begin{figure}[htbp]
    \centering
    \includegraphics[width=0.6\textwidth]{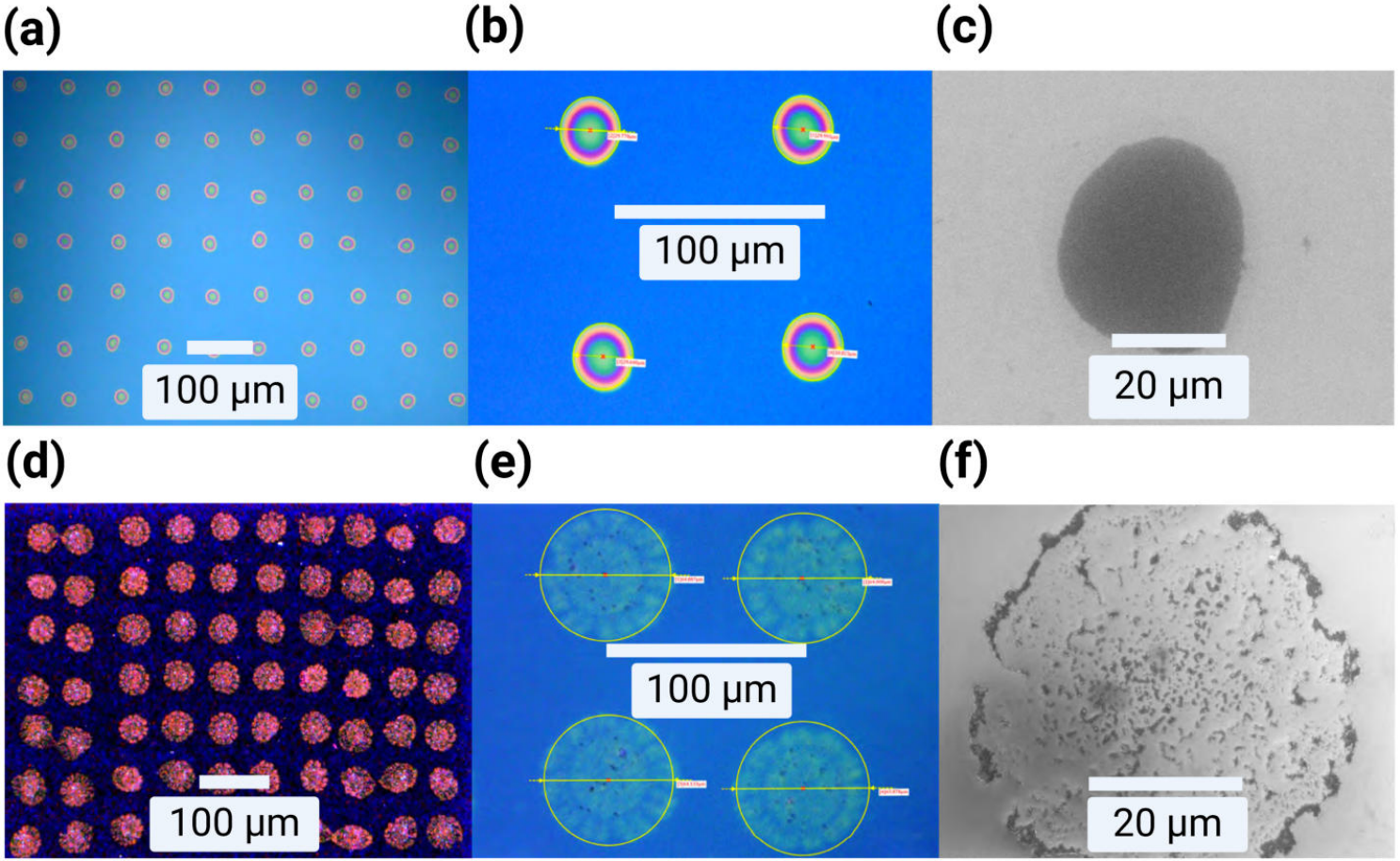}
    \caption{
        \textbf{Surface characterization of inkjet-printed UV and ZIF-8 inks.} 
        (a, b) Optical microscopy images of UV ink drops printed on a SiO$_2$/Si wafer at 300$\times$ and 1500$\times$ magnifications, showing circularity with an average droplet diameter of 29.86 $\mu$m.
        (c) FESEM image of a single UV ink drop, revealing its surface morphology.
        (d, e) Optical microscopy images of ZIF-8 ink drops printed on a SiO$_2$/Si wafer at 300$\times$ and 1500$\times$ magnifications, exhibiting circularity with an average diameter of 65.45 $\mu$m.
        (f) FESEM image of a single ZIF-8 ink drop, highlighting the surface texture.
    }
    \label{fig:inkjet-characterization}
\end{figure}

As shown in Fig.~\ref{fig:inkjet-characterization}, optical microscopy images in (a, b) confirm the uniform distribution and well-defined circularity of UV ink droplets, with an average diameter of 29.86 $\mu$m. The FESEM image (c) provides further insights into the smooth surface morphology of the printed UV ink.
In contrast, inkjet-printed ZIF-8 droplets (Figures~\ref{fig:inkjet-characterization}(d, e)) exhibit a larger average diameter of 65.45 $\mu$m, indicating a different spreading behavior due to variations in ink formulation. The FESEM image (f) shows the rough, porous texture of the printed ZIF-8 ink, confirming its suitability for VOC sensing applications.

\CHANGE{We note that the ring-like contrast visible in Fig.~S10(f) for an isolated ZIF-8 droplet on SiO\textsubscript{2}/Si is consistent with a partial coffee-ring effect during drying. These single-droplet experiments were performed on flat wafers to optimize droplet formation and to study local morphology at high magnification. In the actual sensor arrays, multiple droplets are printed with significant overlap to form continuous films over groups of pixels (see Fig.~5(b–e) in the main text). In this regime, individual coffee rings
merge, and thickness variations are smoothed out at the pixel scale, as reflected by the homogeneous capacitance-response maps across the selected functionalized regions.}

These observations validate the effectiveness of the inkjet-printing process, ensuring well-defined droplet formation and uniform deposition, which is crucial for consistent sensor performance.

\subsection{FESEM Analysis of Printed ZIF-8 Ink Nanoparticles}
\label{subsec:fesem-analysis}

\begin{figure}[htbp]
    \centering
    \includegraphics[width=0.7\textwidth]{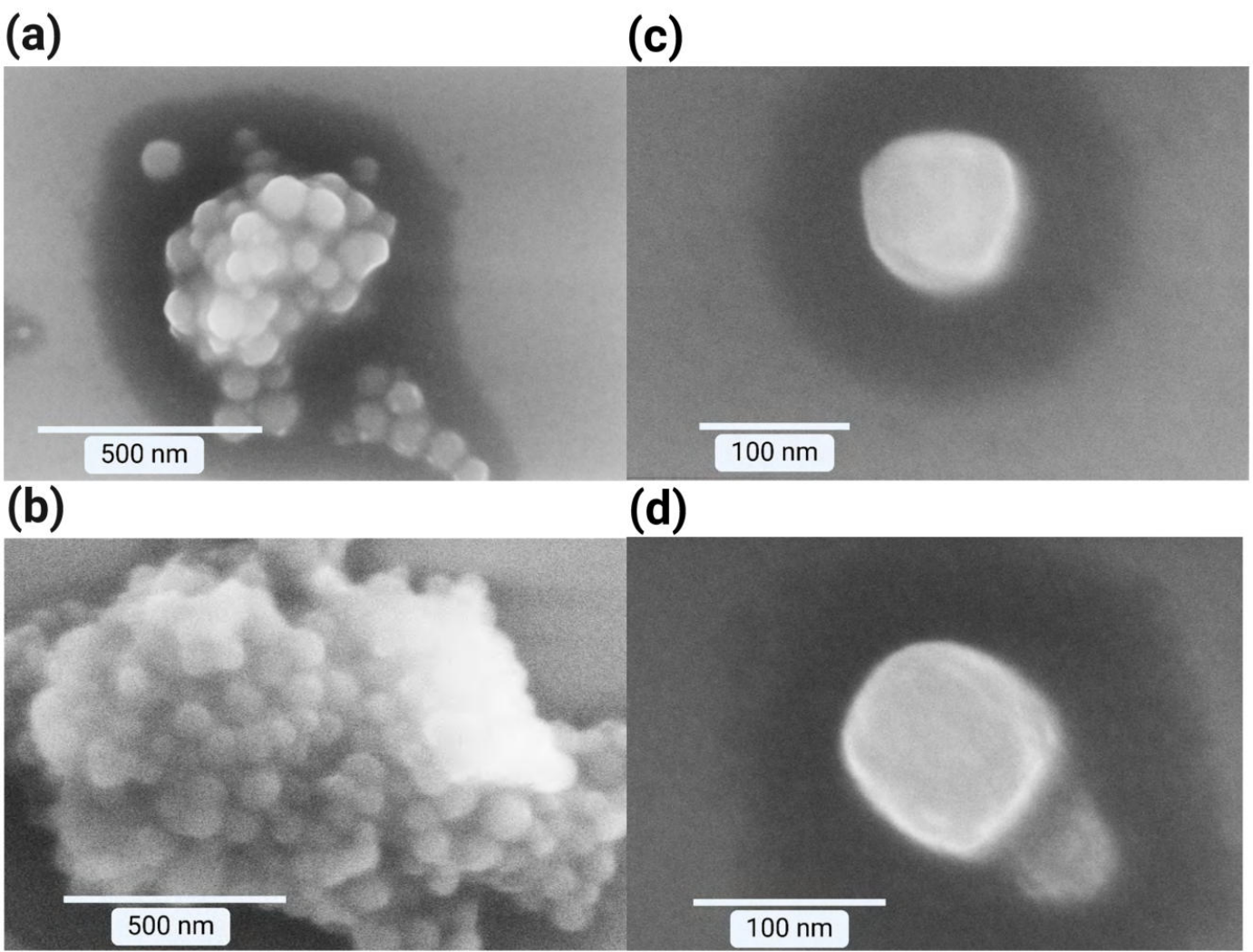}
    \caption{
        \textbf{High-resolution FESEM images of printed ZIF-8 inks.}  
        (a) Aggregated ZIF-8 nanoparticles at 500 nm scale.  
        (b) High-magnification FESEM image of ZIF-8 nanostructures at 500 nm scale.  
        (c) Individual ZIF-8 nanoparticle at 100 nm scale, revealing its morphology.  
        (d) Isolated ZIF-8 nanoparticle with additional nanostructures at 100 nm scale.
    }
    \label{fig:FESEM-ZIF8}
\end{figure}

This section presents high-resolution FESEM images of inkjet-printed ZIF-8 ink on a wafer. Fig.~\ref{fig:FESEM-ZIF8}(a) presents a FESEM image showing aggregated ZIF-8 nanoparticles with a 500 nm scale bar, highlighting the clustered morphology. Fig.~\ref{fig:FESEM-ZIF8}(b) provides a higher-magnification image of ZIF-8 nanostructures at 500 nm scale, revealing the intricate surface morphology. Fig.~\ref{fig:FESEM-ZIF8}(c,d) focus on an individual ZIF-8 nanoparticles. These FESEM analyses show that the inkjet-printed ZIF-8 ink retains the morphology of the synthesized ZIF-8 before undergoing the ink formulation process.

\subsection{EDX Analysis of Inkjet-Printed UV and ZIF-8 Inks}
\label{subsec:edx-analysis}

\begin{figure}[htbp]
    \centering
    \includegraphics[width=0.7\textwidth]{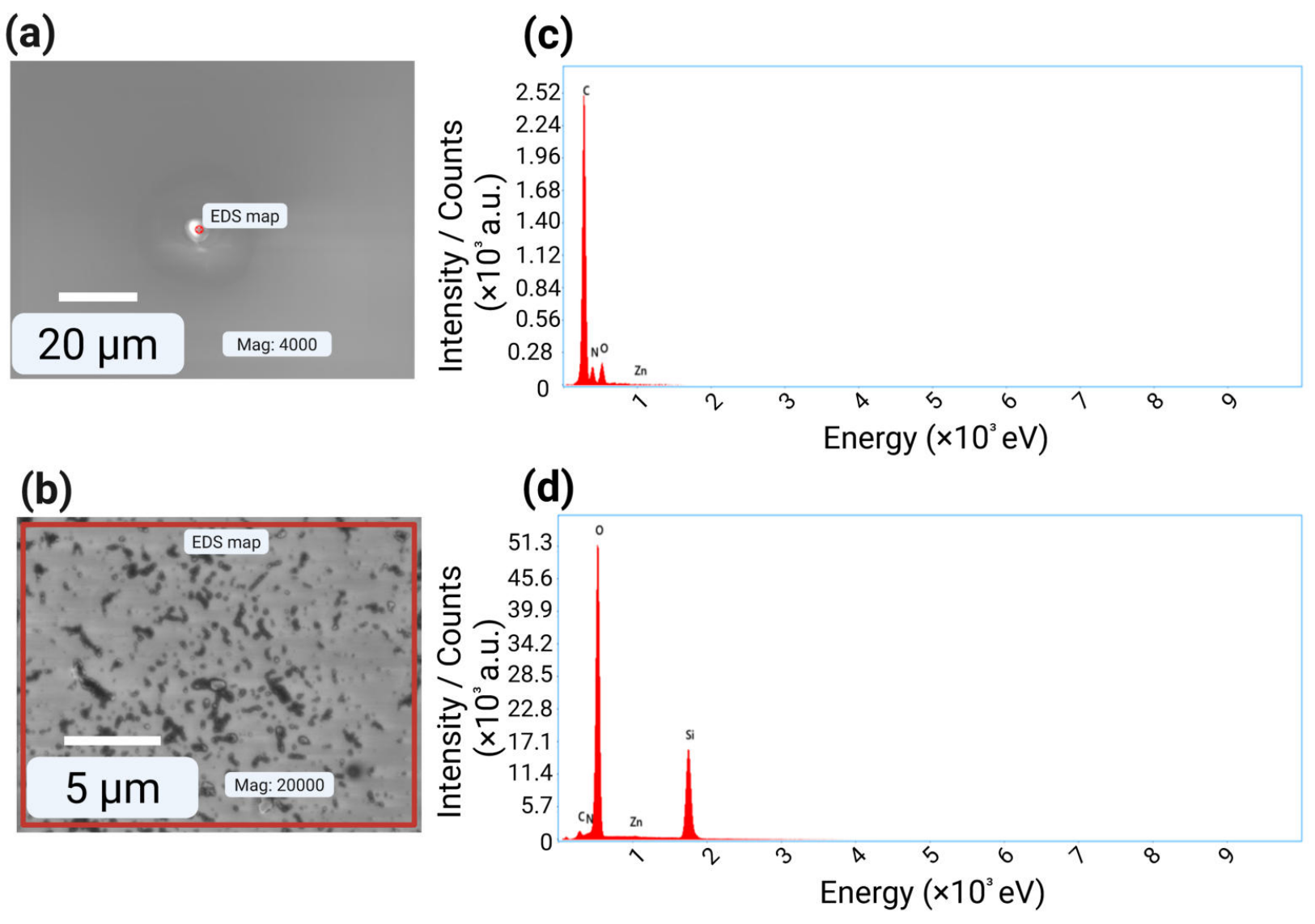}
    \caption{
        \textbf{EDX Analysis of UV and ZIF-8 Inks.}  
        (a) Spot analysis of UV ink at 4000$\times$ magnification.  
        (b) Total EDX intensity-image of ZIF-8 at 20000$\times$ magnification.  
        (c) EDX spectrum of UV ink, highlighting elemental composition.  
        (d) EDX spectrum of ZIF-8 ink, confirming Zn, O, and Si presence.
    }
    \label{fig:EDX-analysis}
\end{figure}

The total EDX intensity–image analysis was conducted to investigate the elemental composition of inkjet-printed UV and ZIF-8 inks. Figure~\ref{fig:EDX-analysis} compiles the SEM panels and corresponding EDX spectra. \textbf{Fig.~\ref{fig:EDX-analysis}(a)} shows the SEM view and spot used for the UV-ink EDX measurement (scale bar: 20\,\(\mu\)m), while \textbf{Fig.~\ref{fig:EDX-analysis}(b)} shows the mapped ZIF-8 region (scale bar: 5\,\(\mu\)m). The associated spectra are plotted in \textbf{Fig.~\ref{fig:EDX-analysis}(c)} (UV ink) and \textbf{Fig.~\ref{fig:EDX-analysis}(d)} (ZIF-8 ink), with the x-axis labeled \emph{Energy} \((\times 10^{3}\ \mathrm{eV})\) and the y-axis labeled \emph{Intensity / Counts} \((\times 10^{3}\ \mathrm{a.u.})\). In \textbf{Fig.~\ref{fig:EDX-analysis}(c)}, peaks for C, N, and O dominate the UV-ink spectrum, with a weak Zn signal. In \textbf{Fig.~\ref{fig:EDX-analysis}(d)}, pronounced Zn lines confirm ZIF-8, and peaks from C, N, O, and Si (substrate) are also observed. \textbf{Tables~\ref{table:EDX-UV} and~\ref{table:EDX-ZIF8}} list the detected elements together with their weight and atomic percentages.

\begin{table}[ht]
    \centering
    \small
    \caption{Elemental composition of inkjet-printed UV ink from EDX analysis.}
    \label{table:EDX-UV}
    \begin{tabular}{|c|c|c|c|c|c|c|c|c|c|}
        \hline
        \textbf{Element} & \textbf{Line} & \textbf{Weight\%} & \textbf{MDL} & \textbf{Atomic\%} & \textbf{Error\%} & \textbf{Net Int.} & \textbf{R} & \textbf{A} & \textbf{F} \\
        \hline
        C  & K & 97.5  & 0.63 & 98.0  & 11.2 & 686.6  & 0.9641  & 0.9700  & 1.0000 \\
        N  & K & 0.9   & 0.87 & 0.8   & 71.0 & 8.0    & 0.9701  & 0.7904  & 1.0000 \\
        O  & K & 1.6   & 0.84 & 1.2   & 37.6 & 16.0   & 0.9750  & 0.8939  & 1.0000 \\
        Zn & L & 0.0   & 0.00 & 0.0   & 100.0 & 0.0   & 0.9875  & 0.9852  & 1.0004 \\
        \hline
    \end{tabular}
\end{table}

\begin{table}[htbp]
    \centering
    \small
    \caption{Elemental composition of inkjet-printed ZIF-8 ink from EDX analysis.}
    \label{table:EDX-ZIF8}
    \begin{tabular}{|c|c|c|c|c|c|c|c|c|c|}
        \hline
        \textbf{Element} & \textbf{Line} & \textbf{Weight\%} & \textbf{MDL} & \textbf{Atomic\%} & \textbf{Error\%} & \textbf{Net Int.} & \textbf{R} & \textbf{A} & \textbf{F} \\
        \hline
        C  & K  & 4.7   & 0.17 & 7.4   & 11.1  & 41.9   & 0.9225  & 0.4555  & 1.0000 \\
        N  & K  & 1.5   & 0.09 & 2.0   & 9.3   & 33.3   & 0.9299  & 0.6500  & 1.0000 \\
        O  & K  & 56.6  & 0.07 & 66.1  & 6.0   & 1877.2 & 0.9359  & 0.7938  & 1.0000 \\
        Si & K  & 36.8  & 0.10 & 24.5  & 6.0   & 760.4  & 0.9663  & 0.9680  & 1.0023 \\
        Zn & L  & 0.3   & 0.16 & 0.1   & 20.6  & 5.1    & 0.9509  & 0.8534  & 1.0025 \\
        \hline
    \end{tabular}
\end{table}

The EDX analysis in Tables~\ref{table:EDX-UV} and~\ref{table:EDX-ZIF8} shows significant differences in elemental composition between the UV ink and ZIF-8 ink.
UV Ink Composition (Table~\ref{table:EDX-UV}): The EDX spectrum confirms that UV ink is mainly composed of carbon (97.5\%), with small traces of nitrogen and oxygen. 
The minimal oxygen presence (1.6\%) suggests the ink remains relatively hydrophobic.

In contrast, the ZIF-8 Ink Composition (Table~\ref{table:EDX-ZIF8}) and elemental distributions in Fig.~\ref{fig:EDX-analysis} show a different characteristic. Partly this is due to the thinner layer thickness (see Fig.~\ref{fig:inkjet-characterization}(f)), such that the substrate elements become more visible. This possibly partly explains the higher silicon peak (Si K = 36.8\%) and the higher oxygen content (56.6\%). The Zn content (0.3\%) indicates the presence of zinc as evidence for ZIF-8 and confirms the successful printing of ZIF-8 ink, as also indicated by the presence of Zn, N, and Si elements, which are characteristic of the MOF. The elemental distributions in Fig.~\ref{fig:EDX-analysis} further validate these observations.

\subsection{Repeatability in Inkjet-Printable UV and ZIF-8 Inks}
\label{sec:repeatability-inks}

\begin{figure}[htbp]
    \centering
    \includegraphics[width=0.7\textwidth]{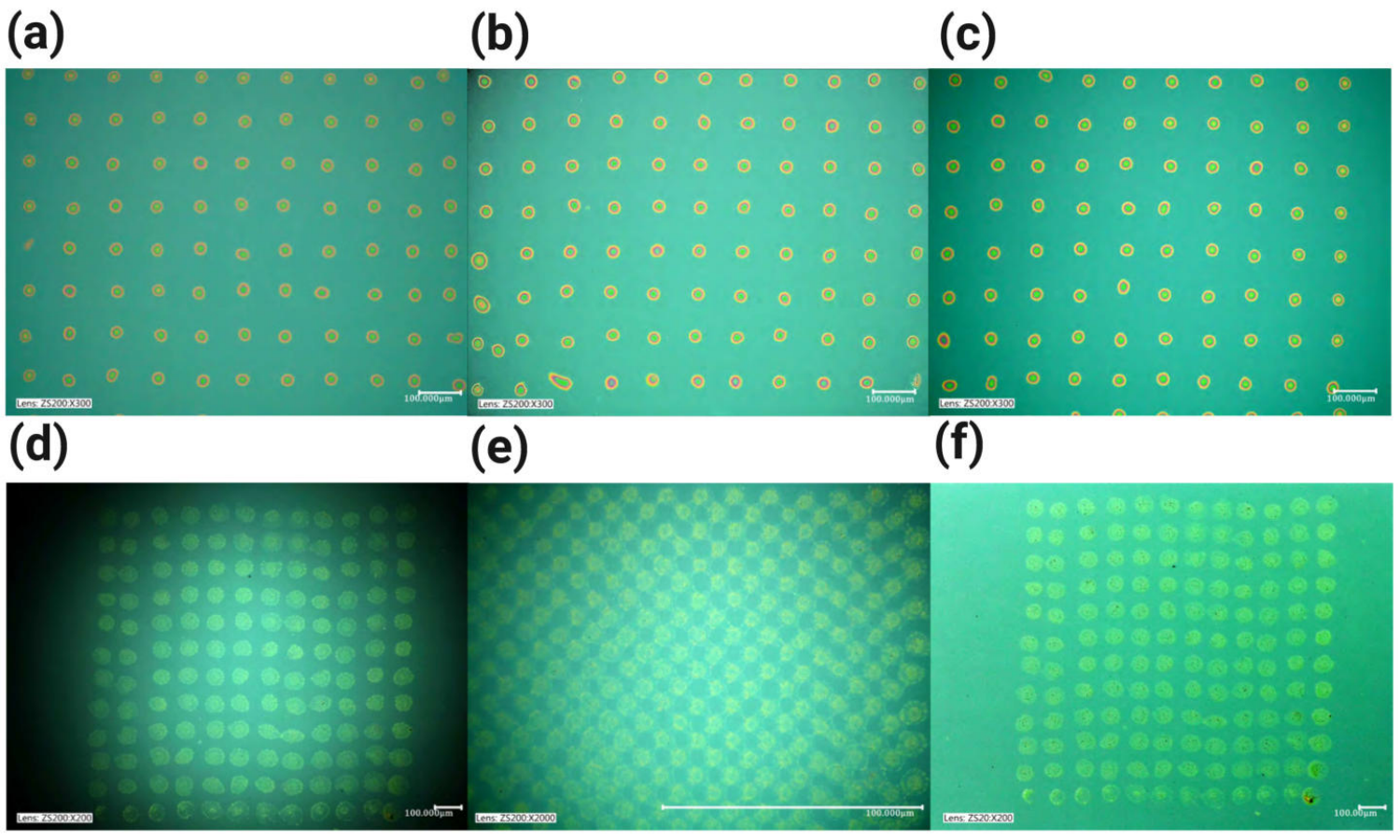}
    \caption{
        \textbf{Optical microscopy images of inkjet-printed UV and ZIF-8 inks.}  
        (a-c) High-resolution images of UV ink patterns printed at 300$\times$ magnification, demonstrating uniform drop distribution.  
        (d-f) Optical microscopy images of ZIF-8 ink patterns under different magnifications, highlighting repeatability and ink uniformity.
    }
    \label{fig:inkjet-patterns}
\end{figure}

The optical microscopy images in Fig.~\ref{fig:inkjet-patterns} illustrate the repeatability and uniformity of the inkjet-printed UV and ZIF-8 inks across multiple printing patterns. 

\textbf{Figures~\ref{fig:inkjet-patterns}(a-c)} show inkjet-printed UV ink drops at 300$\times$ magnification. The drops maintain consistent circularity and even distribution, demonstrating excellent repeatability.
Similarly, \textbf{Figures~\ref{fig:inkjet-patterns}(d-f)} represent ZIF-8 ink patterns under different magnifications. The ink successfully forms an array of uniform spots, confirming stability in ink formulation and droplet deposition. The high consistency in both UV and ZIF-8 ink patterns suggests the functionalization process by inkjet printing can be suitable for high-volume fabrication flows, where repeatability and precision are essential.

\section{Experimental Gas Sensing Setup}
\label{subsec:gas-sensing-setup}

\begin{figure}[htbp]
    \centering
    \includegraphics[width=0.7\textwidth]{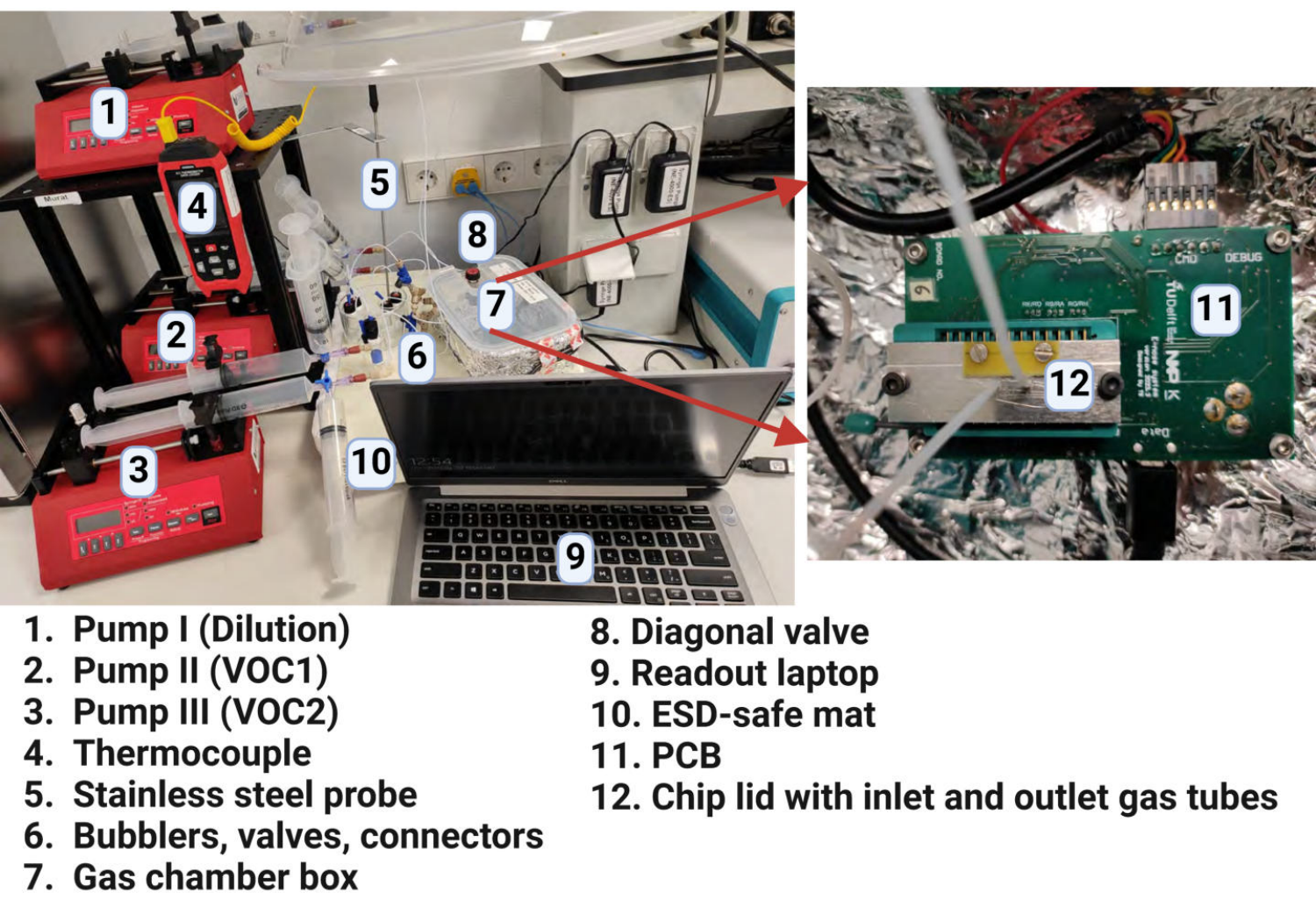}
    \caption{
        Experimental gas sensing setup.  
        The system includes (1) Pump I for dilution, (2) Pump II for VOC1, and (3) Pump III for VOC2 to control gas flow rates. A thermocouple (4) monitors the temperature, while bubblers, valves, and connectors (6) regulate gas flow. The gas chamber box (7) houses the sensing chip, with a diagonal valve (8) enabling precise gas switching. Sensor data is collected via the readout laptop (9) connected to the PCB (11), which processes the signals from the sensor chip. The chip is enclosed under a lid with inlet and outlet gas tubes (12) for controlled exposure.
    }
    \label{fig:lab-setup}
\end{figure}

To conduct controlled gas sensing experiments, a custom-built experimental setup was developed as schematically shown in the main manuscript, while the real picture of the setup is shown in Fig.~\ref{fig:lab-setup}. The system consists of three syringe pumps: Pump I for dilution, Pump II for VOC1, and Pump III for VOC2, which precisely control nitrogen and VOC flow rates. A thermocouple (4) is used to monitor the temperature during experiments to ensure stable vapor pressure conditions. 
The gas mixing and delivery system includes bubblers, valves, and connectors (6) to regulate VOC saturation and avoid condensation before exposure to the sensor. The gas mixture is directed into a gas chamber box (7), where the PCB and chip are placed. A diagonal valve (8) is integrated into the system, allowing for controlled switching between nitrogen and VOC exposure cycles. 
The sensor response is recorded using a readout laptop (9), which interfaces with the PCB (11) that processes the sensor signals. The sensing chip is housed under a chip lid with inlet and outlet gas tubes (12), ensuring precise gas delivery. The entire system operates on an ESD-safe mat (10) to prevent electrostatic discharge from affecting the measurements.
In the next subsections, we will describe the capacitive electrode structure, the readout electronics, and the data analysis.

\subsection{Microelectrode Structure and Dimensions}
\label{subsec:microelectrode-structure}

\begin{figure}[htbp]
    \centering
    \includegraphics[width=0.6\textwidth]{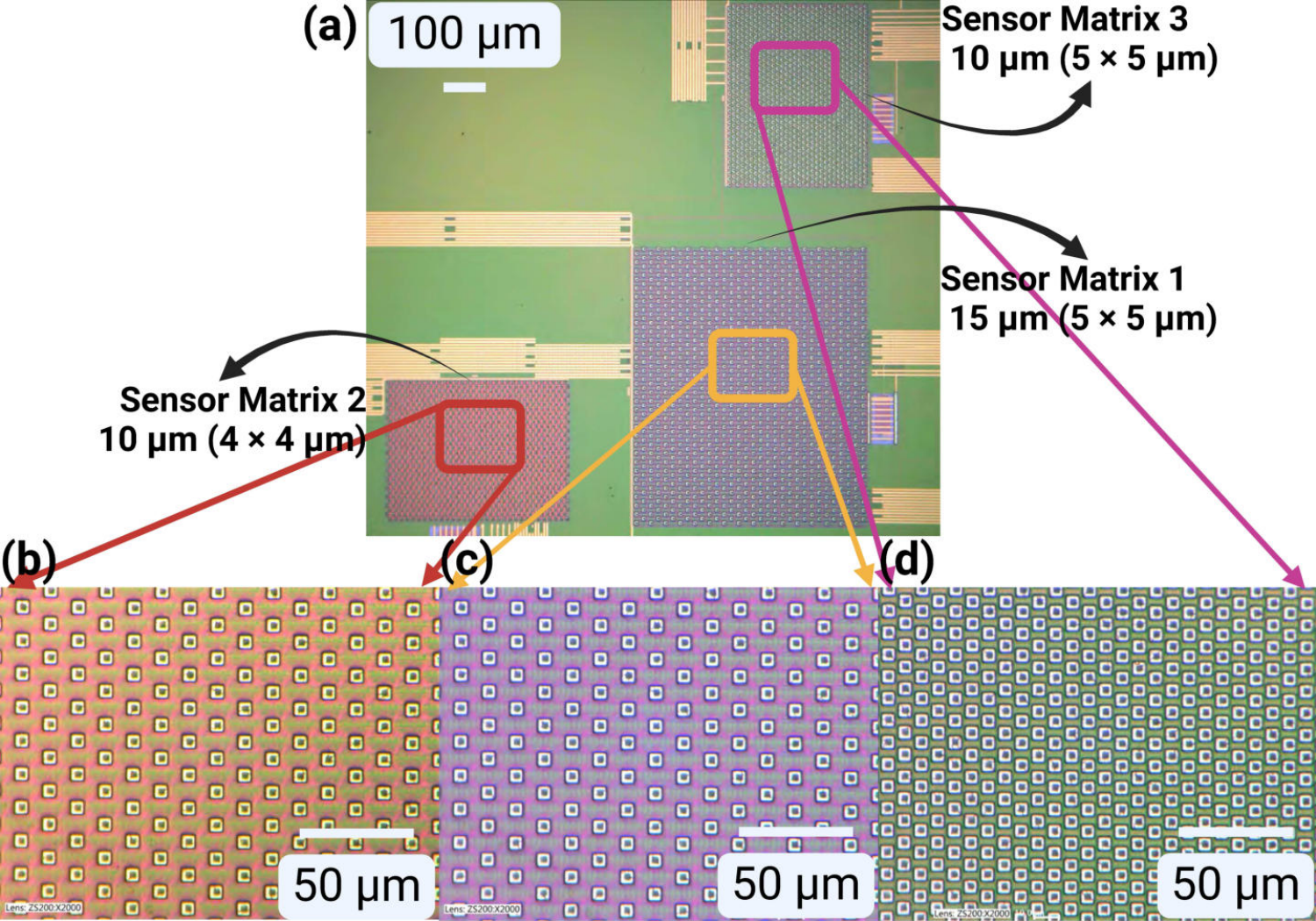}
    \caption{
        \textbf{CMOS microelectrode structures}. 
        (a) Overview of the CMOS chip layout with three sensor matrices. Each matrix has 32 columns and 32 rows of electrodes connected by the sensor readout interface circuit.
        (b) Magnified view of Sensor Matrix 2, featuring 4 $\times$ 4 $\mu$m electrodes with 10 $\mu$m interelectrode spacing.  
        (c) Magnified view of Sensor Matrix 3, featuring 5 $\times$ 5 $\mu$m electrodes with 10 $\mu$m interelectrode spacing.  
        (d) Magnified view of Sensor Matrix 1, featuring 5 $\times$ 5 $\mu$m electrodes with 15 $\mu$m interelectrode spacing.
    }
    \label{fig:chip-microelectrodes}
\end{figure}

To better understand the pixelated sensing architecture of the CMOS chip, Fig.~\ref{fig:chip-microelectrodes} presents a detailed optical characterization of the microelectrode structures. The chip consists of multiple sensor matrices with varying electrode sizes and spacing, which play a critical role in influencing the ink distribution and capacitive response.

Fig.~\ref{fig:chip-microelectrodes}(a) provides an overview of the CMOS chip layout, highlighting three distinct sensor matrices. The magnified images in Figures~\ref{fig:chip-microelectrodes}(b-d) illustrate the arrangement and spacing of the microelectrode arrays.

\subsection{Derivation of Capacitance Measurement from ADC Output}
\label{sec:capacitance-derivation}

The relation between the readout ADC value and the capacitance $C$ of the sensor is explained in this section.  

In the employed switched capacitor circuit, the measured current \( I_{\text{measured}} \) is related to the capacitance by:

\begin{equation}
    I_{\text{measured}} = C \cdot V \cdot f
\end{equation}

Where:
\begin{itemize}
    \item \( C \) is the measured capacitance.
    \item \( I_{\text{measured}} \) is the current flow from the switched capacitor.
    \item \( V \) is the capacitor voltage when charged.
    \item \( f \) is the charge/discharge clock frequency.
\end{itemize}

For this specific measurement system:

\begin{equation}
    I_{\text{measured}} = C \cdot 0.9\,V \cdot 40\,\text{MHz}
\end{equation}

A current mirror amplifies the current by a factor \( n \), given by:

\begin{equation}
    I_{\text{read}} = n \cdot I_{\text{measured}}
\end{equation}

Where:
\begin{itemize}
    \item \( I_{\text{read}} \) is the output current from the current mirror.
    \item \( n \) is the current magnification factor of the mirror, given as \( n = 100 \).
\end{itemize}

Thus, substituting \( n = 100 \):

\begin{equation}
    I_{\text{read}} = 100 \cdot I_{\text{measured}}
\end{equation}

The current \( I_{\text{read}} \) is converted into a voltage using a transimpedance amplifier:

\begin{equation}
    V_{\text{measured}} = V_{\text{bias}} + R \cdot I_{\text{read}}
\end{equation}

Where:
\begin{itemize}
    \item \( V_{\text{measured}} \) is the output voltage of the transimpedance amplifier.
    \item \( V_{\text{bias}} = 0.5\,V \) is the bias voltage for the operational amplifier.
    \item \( R = 68\,k\Omega \) is the feedback resistor in the amplifier circuit.
\end{itemize}

Substituting the values:

\begin{equation}
    V_{\text{measured}} = 0.5\,V + 68\,k\Omega \cdot I_{\text{read}}
\end{equation}

The voltage is converted into a digital value by the ADC using:

\begin{equation}
    \text{ADC}_{\text{code}} = \frac{V_{\text{measured}}}{V_{\text{max}}} \times \text{ADC}_{\text{range}}
\end{equation}

Where:
\begin{itemize}
    \item \( V_{\text{max}} = 3.3\,V \) is the maximum voltage the ADC can read.
    \item \( \text{ADC}_{\text{range}} = 4096 \) is the maximum ADC output value (corresponding to a 12-bit ADC).
    \item \( \text{ADC}_{\text{code}} \) (or \( V_{\text{read}} \)) is the digital value observed from the ADC.
\end{itemize}

Substituting values:

\begin{equation}
    \text{ADC}_{\text{code}} = \frac{V_{\text{measured}}}{3.3\,V} \times 4096
\end{equation}

Rearranging all the previous equations and solving for \( C \):

\begin{equation}
    C_{\text{measured}} = \frac{1}{100 \times 0.9\,V \times 40\,\text{MHz} \times 68\,k\Omega} \times 
    \left[ \frac{(\text{ADC}_{\text{output}} - 620) \times 3.3\,V}{4096} \right]
\end{equation}

This final equation provides a direct method to calculate the measured capacitance from the ADC output.

\subsection{Data Analysis of Pixelated Capacitive Response}
\label{subsec:pixelated-capacitive-analysis}

\begin{figure}[htbp]
    \centering
    \includegraphics[width=0.7\textwidth]{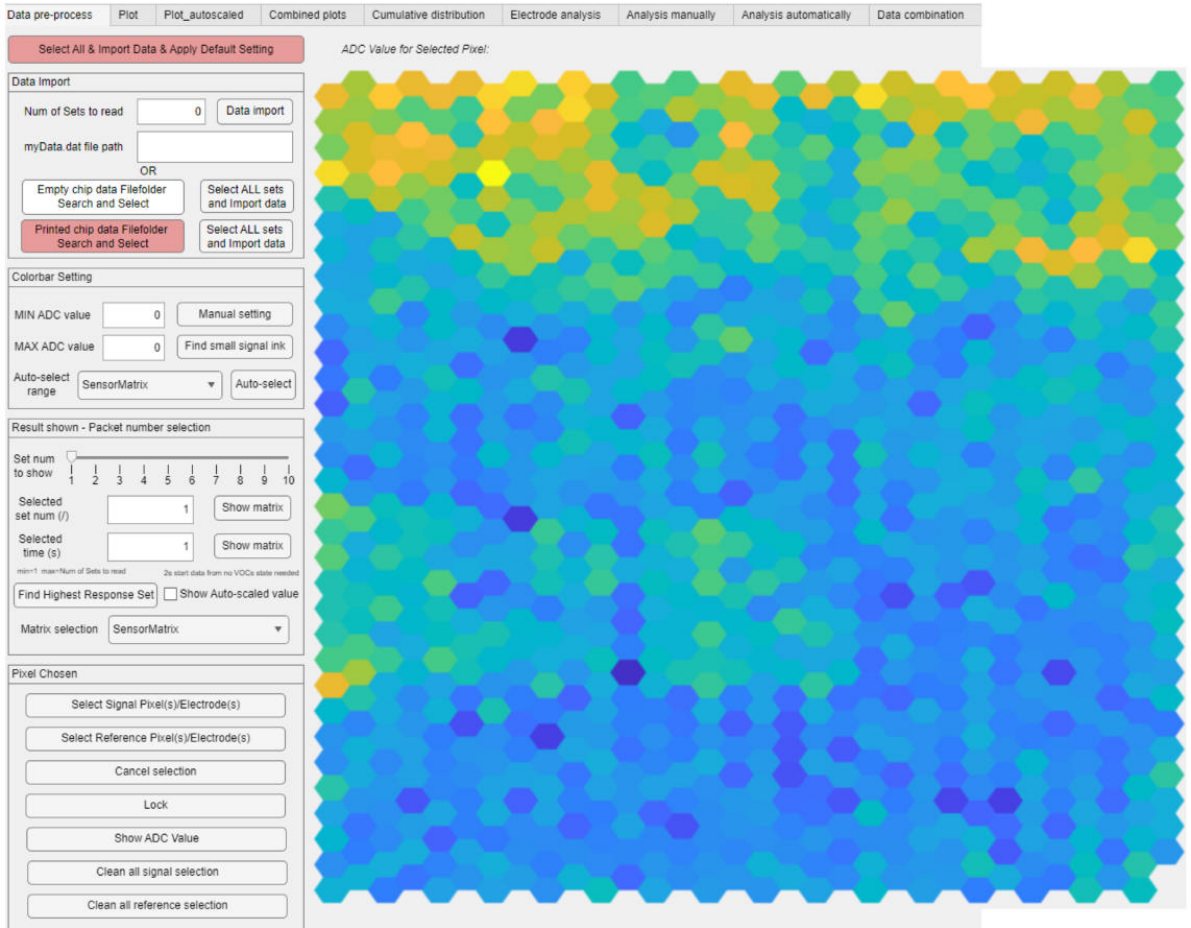}
    \caption{
        \textbf{Software for Data Analysis of Pixelated Capacitive Response.}  
        The developed software allows users to select and analyze capacitive responses from ink-coated and non-coated sensor areas. This ensures precise measurement and differentiation of signal variations across the array. The selection tool enables defining regions of interest (ROI). In the current study, we averaged multiple pixels to reduce the effects of noise.
    }
    \label{fig:data-analysis}
\end{figure}

A custom-built software was developed to process and analyze the output of all 1024 pixelated capacitive response sensors in a matrix. This software enables the selection of ink-coated and non-coated areas, allowing average ADC values and differences between average values like $\Delta \text{ADC}_{\text{ink, VOC}}$ to be plotted as a function of time. Fig.~\ref{fig:data-analysis} illustrates the software interface, where a pixelated capacitive heatmap is displayed, allowing users to visualize signal variations across the sensor array and select the different relevant functionalized regions on the matrix for analysis.

The software provides a heatmap visualization of sensor responses, as shown in Fig.~\ref{fig:data-analysis}, which aids in detecting and comparing sensor signals across different regions of the chip. This capability is crucial for ensuring reproducibility and accuracy in VOC sensing applications. Furthermore, an average response was calculated over multiple selected pixels to reduce noise, ensuring a more reliable and representative dataset. This methodology helps compensate for potential non-uniformities in ink deposition during the inkjet printing process.

\label{sec:voc-water-sensing-mechanism}

\begin{table}[htb]
    \centering
    \small
    \renewcommand{\arraystretch}{1.3} 
    \caption{
        List of materials used in this study, including their dielectric constants, solubility interactions, ligand types, cluster types, pore dimensions, largest included sphere diameters (\(d_i\)), largest free sphere diameters (\(d_f\)), and crystalline densities (\(\rho_c\)).
        DC: Dielectric constant, SI: Solubility interactions, 
        \(d_i\): Largest included sphere diameter (\r{A}), \(d_f\): Largest free sphere diameter (\r{A}), 
        \(\rho_c\): Crystalline density (g/mL).
        Ligands: TPA = Terephthalate, mIm = Methyl-imidazolate.
        cus = Coordinatively unsaturated site.
    }
    \label{tab:materials}
    \resizebox{\textwidth}{!}{ 
    \begin{tabular}{l p{2.5cm} p{3cm} c p{3.5cm} c c c c}
        \toprule
        \textbf{Materials} & \textbf{Ligand} & \textbf{Cluster Type} & \textbf{DC} & \textbf{SI} & \textbf{Pore Dim.} & \textbf{\(d_i\) (\r{A})} & \textbf{\(d_f\) (\r{A})} & \textbf{\(\rho_c\) (g/mL)} \\
        \midrule
        \textbf{UV ink} & N/A & N/A & 2.75 \cite{zhang2023low} & Dipolarity, Dispersion Forces \cite{grate1991solubility} & N/A & N/A & N/A & 1.03--1.06 \cite{amcadgraphics} \\
        \textbf{ZIF-8} & mIm & Zn & 1.1--1.5 \cite{ryder2018dielectric} & Porous adsorption, Hydrophobic Interactions \cite{somphon2023zif} & 3D & 11.4 & 3.4 & 0.92 \cite{park2006exceptional}\\
        \textbf{MIL-101} & TPA & Cr\textsubscript{3}O(X) (cus)\textsubscript{2} & N/A  & Porous adsorption, Dipolarity \cite{navid2025calix, zou2022advances} & 3D & 33.7 & 13.9 & 0.48 \cite{ferey2005chromium}\\
        \textbf{MIL-140A} & TPA & [Zr\textsubscript{2}O]\textsubscript{n} & N/A & Porous adsorption, Dipolarity \cite{schulz2019low} & 1D & 3.9 & 3.1 & 1.69 \cite{guillerm2012series}\\
        \bottomrule
    \end{tabular}
    }
\end{table}

\section{Sensor Response to Mixed toluene and 2-butanone at Constant Total Concentration}
\label{sec:mix_response}

\begin{figure}[htbp]
    \centering
    \includegraphics[width=0.50\linewidth]{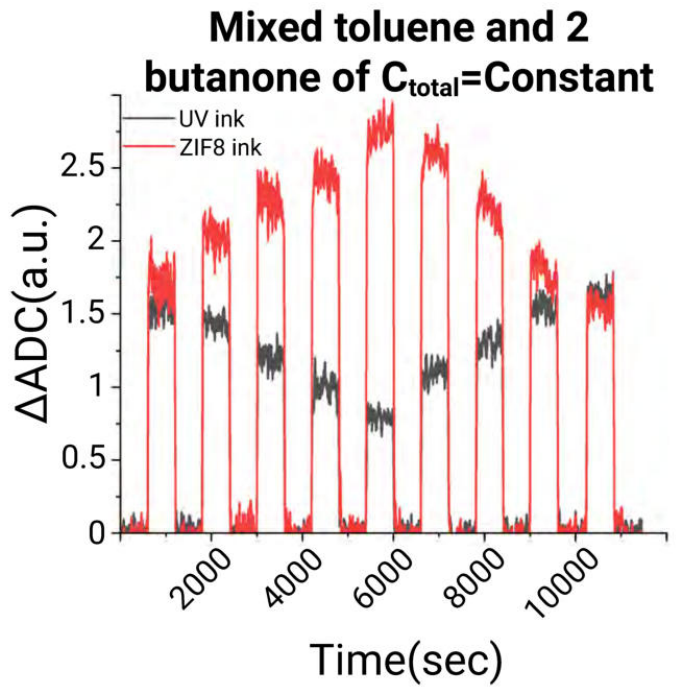}
    \caption{\CHANGE{\textbf{Dynamic response of UV and ZIF-8 inks to stepped binary mixtures at fixed total VOC load.} 
    The sensor is exposed to mixtures of toluene and 2-butanone while the total VOC concentration is kept constant at 
    $C_{\mathrm{total}} = 12{,}000$~ppm. The mixing ratio is varied stepwise, and the resulting 
    $\Delta$ADC responses of the UV ink and ZIF-8 ink are monitored over time.}}
    \label{fig:mix_dynamic}
\end{figure}

\CHANGE{Fig.~\ref{fig:mix_dynamic} shows a stepped-ratio experiment in which $C_{\text{toluene}}$ and 
$C_{\text{2-butanone}}$ were alternated while the total VOC level remained fixed at 
$C_{\text{toluene}}+C_{\text{2-butanone}}=12{,}000$~ppm during the following nine mixing steps:}
\[
\begin{aligned}
C_{\text{toluene}} &= \{10{,}000,\; 8{,}000,\; 6{,}000,\; 4{,}000,\; 2{,}000,\\[-2pt]
&\qquad 4{,}000,\; 6{,}000,\; 8{,}000,\; 10{,}000\}\,\text{ppm},\\[6pt]
C_{\text{2-butanone}} &= \{2{,}000,\; 4{,}000,\; 6{,}000,\; 8{,}000,\; 10{,}000,\\[-2pt]
&\qquad 8{,}000,\; 6{,}000,\; 4{,}000,\; 2{,}000\}\,\text{ppm}.
\end{aligned}
\]

\CHANGE{The $\Delta$ADC responses of the UV and ZIF-8 functionalized regions clearly follow opposite trends as the
mixture composition is stepped: when the toluene fraction increases (and 2-butanone decreases), the UV-ink signal 
rises while the ZIF-8 signal drops, and vice versa. This behavior confirms that the paired sensor responses carry 
information about the relative composition of non-equimolar binary mixtures, even when the total VOC concentration 
is constant. In other words, the array can resolve composition changes rather than merely tracking the overall VOC 
load.}

\CHANGE{In the present work, the mixture experiment is used qualitatively to demonstrate this complementary behavior.
A full quantitative decoupling of the two analytes would require collecting a systematic calibration data set over 
a grid of $(C_{\text{toluene}}, C_{\text{2-butanone}})$ combinations and fitting an appropriate multivariate model. 
Designing such an extended calibration protocol and upgrading the gas-mixing manifold for more flexible ratio control 
are left for future work, but the results in Fig.~\ref{fig:mix_dynamic} already illustrate the feasibility of 
simultaneous multi-component sensing on a single CMOS capacitive chip.}

\section{Limit of Detection (LOD) Calculation}
\label{sec:LOD}

\CHANGE{To estimate the theoretical limit of detection (LOD) for each ink--VOC combination,
we used the conventional $3\sigma$ criterion based on the baseline noise of the
$\Delta\mathrm{ADC}$ signal and the slope of the calibration curve ~\cite{luo2023chemiresistive}.}

\CHANGE{First, we calculated the root-mean-square (rms) noise of the baseline prior to VOC
exposure. For each functional region (UV ink and ZIF-8 ink), we selected a baseline
segment in dry N$_2$ before the analyte step, extracted $N$ consecutive data points
of the averaged $\Delta\mathrm{ADC}$ signal, and fitted the baseline with a low-order
polynomial to account for slow drift. The residual sum of squares $V_{\chi^2}$ was
computed as}
\begin{equation}
  V_{\chi^2} = \sum_{i=1}^{N} \bigl(y_i - y\bigr)^2,
\end{equation}
\CHANGE{where $y_i$ is the measured $\Delta\mathrm{ADC}$ at time point $i$ and $y$ is the
corresponding value from the baseline fit. The rms baseline noise was then obtained as}
\begin{equation}
  \mathrm{rms}_{\mathrm{noise}} = \sqrt{\frac{V_{\chi^2}}{N}}.
\end{equation}
\CHANGE{For the averaged responses reported in Figs.~7(a–d), this procedure yielded
$\mathrm{rms}_{\mathrm{noise}} \approx 0.03$~ADC codes.
Next, we determined the sensitivity $S$ from the calibration curves in
Figs.~7(e,f). For ink--VOC pairs that follow a linear
relationship,}

\begin{equation}
  \Delta\mathrm{ADC}_{\mathrm{ink,VOC}} = c_{\mathrm{ink,VOC},1}\, C_{\mathrm{VOC}},
\end{equation}
\CHANGE{we took the slope $S = c_{\mathrm{ink,VOC},1}$ (ADC/ppm), using the fit parameters
listed in Table~3 of the main text. For the ZIF-8 response
to 2-butanone, which exhibits saturation and was fitted with a quadratic model,}
\begin{equation}
  \Delta\mathrm{ADC}_{\mathrm{ZIF,Bu}} =
    c_{\mathrm{ZIF,Bu},1}\, C_{\mathrm{Bu}} +
    c_{\mathrm{ZIF,Bu},2}\, C_{\mathrm{Bu}}^2,
\end{equation}
\CHANGE{the low-concentration sensitivity was approximated by the initial slope at}
$C_{\mathrm{Bu}}\to 0$, i.e.\ $S \approx c_{\mathrm{ZIF,Bu},1}$.

\CHANGE{Finally, the LOD for each ink--VOC combination was estimated according to}
\begin{equation}
  \mathrm{LOD} = \frac{3\,\mathrm{rms}_{\mathrm{noise}}}{S},
\end{equation}
\CHANGE{with $\mathrm{rms}_{\mathrm{noise}}$ expressed in ADC units and $S$ in ADC/ppm, so that
$\mathrm{LOD}$ is obtained in ppm. Using $\mathrm{rms}_{\mathrm{noise}} \approx 0.03$~ADC
and the slopes from Table~3, the resulting values are summarized in~ Table~\ref{tab:LOD}}

\begin{table}[t]
  \centering
  \caption{Estimated limits of detection (LOD) for different ink--VOC combinations
  based on the $3\sigma$ criterion, using $\mathrm{rms}_{\mathrm{noise}} \approx 0.03$~ADC.}
  \label{tab:LOD}
  \begin{tabular}{lccc}
    \hline
    Ink   & VOC          & Slope $S$ (ADC/ppm) & LOD (ppm) \\
    \hline
    ZIF-8 & 2-butanone   & $1.238\times10^{-3}$ & $\approx 7.3\times10^{1}$ \\
    ZIF-8 & toluene      & $1.830\times10^{-5}$ & $\approx 4.9\times10^{3}$ \\
    UV    & 2-butanone   & $1.031\times10^{-4}$ & $\approx 8.7\times10^{2}$ \\
    UV    & toluene      & $1.153\times10^{-4}$ & $\approx 7.8\times10^{2}$ \\
    \hline
  \end{tabular}
\end{table}

\newpage
\null\newpage
   
\setcounter{page}{33}
\renewcommand{\thepage}{\arabic{page}}

\newpage
\bibliography{sources}

\end{document}